\documentclass[]{tCPH2ep}

\usepackage{psfrag}
\usepackage{color}
\usepackage{axodraw4j}
\usepackage{subfigure}

\renewcommand\({\left(}
\renewcommand\){\right)}

\newcommand{\be}{\begin{equation}}
\newcommand{\ee}{\end{equation}}
\def\bea{\begin{eqnarray}}
\def\eea{\end{eqnarray}}

\newcommand{\base}{\baselineskip 16pt}

\newcommand{\muu}{ m_{\gamma^\prime}}

\newcommand{\MF}{B}

\begin{document}


\markboth{J.~Redondo and A.~Ringwald}{Light shining through walls}


\title{
\vspace{-3cm}
{\small\rm \hfill{DESY 10-175, MPP-2010-149}}\\
\vspace{3cm}
Light shining through walls}

\author{
Javier Redondo$^{a,b}$
and Andreas Ringwald$^{a}$$^{\ast}$
\thanks{$^\ast$Corresponding author. Email: andreas.ringwald@desy.de \vspace{6pt}} 	\\	\vspace{6pt}
$^{a}${\em{Deutsches Elektronen-Synchrotron DESY, Notkestra\ss e 85, D-22607 Hamburg, Germany}}\\
$^{b}${\em{Max-Planck-Institut f\"ur Physik, F\"ohringer Ring 6, D-80805 M\"unchen, Germany}} 	\\	\vspace{6pt}
\received{October 2010}}

\maketitle

\begin{abstract}
Shining light through walls?
At first glance this sounds crazy. However, very feeble gravitational and electroweak effects allow for
this exotic possibility. Unfortunately, with present and near future technologies the opportunity to observe
light shining through walls via these effects is completely out of question.
Nevertheless there are quite a number of experimental collaborations around the globe involved in this quest.
Why are they doing it?
Are there additional ways of sending photons through opaque matter?
Indeed, various extensions of the standard model of particle physics predict the existence of
new particles called WISPs - extremely weakly interacting slim particles.
Photons can convert into these hypothetical particles,  which have no problems to penetrate very dense
materials, and these can reconvert into photons after their passage - as if light was effectively traversing walls.
We review this exciting field of research, describing the most important WISPs, the present and future experiments, the indirect hints from astrophysics and cosmology pointing to the existence of WISPs, and finally outlining the consequences that the discovery of WISPs would have.

\bigskip

\bigskip
\bigskip

\end{abstract}

\section{Introduction}
The idea of making light traverse a wall may seem either completely trivial or
completely impossible. The conclusion depends on what we understand as light and as a
wall. If we understand light in a broad sense as electromagnetic waves, then it is
clear that shining light through walls is trivial: we are used to use our cellular phones
and to listen radio inside buildings thanks to long-wavelength electromagnetic radiation.

If we restrict ourselves to visible, or shorter wavelength electromagnetic radiation,
the situation changes substantially. Of course, there are types of walls that allow light to
propagate across them -- windows are an every-day life example -- but other apparently
opaque bodies (like us) can be transparent to certain wavelengths as, e.g. X-rays.

Electromagnetic radiation or its quanta, the photons, interact quite strongly with the electrons of the wall, being
scattered and absorbed. We could say that the obstacle to achieve light shining through walls is that
the electromagnetic interactions are too strong.

However, this is not the end of the story.
There are in nature other forms of radiation which are far less strongly interacting.
If one could convert photons into quanta of these other forms of radiation, the latter would do
the dirty work of traversing the wall, and all one has to do is to revert the conversion process at
the end of the wall.
In the standard model (SM) of particle physics (for a short review, see Ref.~\cite{Amsler:2008zzb}),
there are two such candidates for this intermediate step,
the gravitons\footnote[1]{Gravitons, i.e. particle-like excitations of the gravitational field, have never been observed as such, albeit there is little or no doubt about their existence.} -- quanta of the gravitational field, which feel only the weakest force in the SM
-- gravity -- and neutrinos, which participate only in the weak interactions
-- the next to weakest known fundamental force.

\begin{figure}[tbp] 
\centering
\includegraphics[width=14cm]{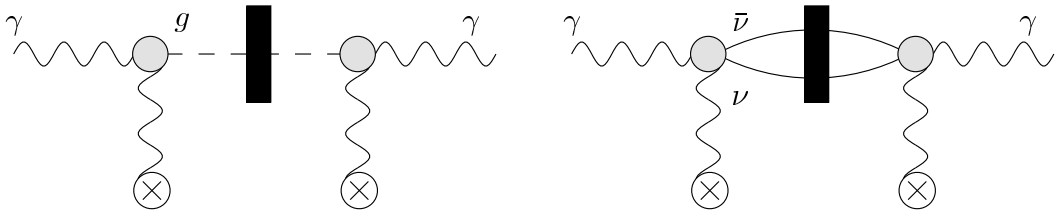}
\caption{In the standard model, light shining through walls can happen via conversion
of photons ($\gamma$) into gravitons ($g$) or neutrino-antineutrino pairs ($\nu,\bar \nu$) in the background of a magnetic field
(marked by a cross).}
\label{LSWsm}
\end{figure}

But can photons convert into gravitons or neutrinos? How can this happen?
A first requirement is that the rest mass of the particle is smaller than the photon energy, otherwise the conservation
of energy would be violated in the conversion.
Photons and gravitons have extremely small masses  -- indeed so small that they have not been measured
yet\footnote[2]{Neutrino oscillation experiments are sensitive to the difference of neutrino squared masses but not to their absolute value. The current data reveals that at least two of the 3 standard neutrinos are massive, the heaviest having a mass above 
$0.06$ eV. The stringest upper limit comes from cosmology which sets the sum of the three neutrino masses to be smaller 
than $\sim 0.5$~eV.
Finally, there are good reasons to believe that gravitons are exactly massless.} --
so they satisfy this criterion.
Another conservation law that has to be satisfied, is angular momentum.
Photons are particles of spin 1, but gravitons have spin 2 and neutrinos 1/2 so the direct conversion
is not possible.
In principle one could combine two photons into a graviton or one photon into two neutrinos but this procedure is typically not optimal.
A practical alternative is to provide the missing angular momentum by an external background field,
for instance a strong and constant magnetic field, pointing in a direction transverse to the photon
velocity\footnote[3]{If it points along the velocity it commutes with the helicity operator and cannot mediate transitions
between particles with different helicities.}.
Therefore, in the presence of a background magnetic field,  light shining through walls is possible
within the SM. Schematic diagrams of these processes are shown in Fig.~\ref{LSWsm}.

Unfortunately, the probabilities of these processes are ridiculously small.
For instance, the probability of shining a photon through a wall via an intermediate graviton
is\footnote[4]{This formula will be understood later in this review.}
\begin{equation}
P(\gamma \to g \to \gamma) \simeq 10^{-83} \(\frac{B}{{\rm T}}\)^4 \(\frac{L}{{\rm m}}\)^4,
\end{equation}
where $B$ is the magnetic field strength and $L$ its length~\cite{Raffelt:1987im}.
State of the art magnets have magnetic fields of few Tesla and lengths of several meters.
The Sun has emitted around $10^{63}$ photons in its entire life so none of them
had gone through a wall. To get a brilliant enough source of photons to
observe such an effect we would have to be really ambitious.
Only if we could gather every photon remaining from the big bang in the visible universe,
one could see some dozens of photons shining through the wall (!).
Clearly the gravitational interactions are too weak.

Neutrinos only interact with the rest of the world through the weak force.
From a fundamental level, the weak force is not so weak however.
As was shown by Glashow, Weinberg, and Salam in the sixties the weak force is in many
senses equivalent to electromagnetism. Indeed, they both are two aspects of a unified
concept: the electroweak force.
Similar to the photon, which mediates the
electromagnetic force, there are particles that mediate the weak forces.
These particles are called $W^+, W^-$ and $Z$ bosons.
The weak forces are indeed so weak because -- for a still unknown reason -- the $W$ and $Z$
bosons are very massive while the mass of the photon is zero.

The masses of the weak bosons are much larger than almost any energy that can be
gathered in a collision of elementary particles in the whole universe -- with the exception
of large particle colliders
and ultra-high energy cosmic rays.
In a classical world these particles could not mediate any kind of interaction because
there would be simply not enough energy to produce them.
However, in the quantum world in which we are living, the uncertainty principle does allow
the creation and propagation of these particles, even if only for very short times.
The small lifetimes of these particles are related to their Compton wavelengths which in turn are
determined by their mass.
Neutrinos are therefore very weakly interacting particles, not because of their interaction strength, but
because the mediators have huge masses and can mediate interactions during only very small times.
Put it another way, using the uncertainty principle in its momentum vs distance form,
neutrinos can interact strongly, but only with particles that are as close as the Compton wavelength of
the $W$ and $Z$ bosons.

\begin{figure}[tbp] 
\centering
\includegraphics[width=7cm]{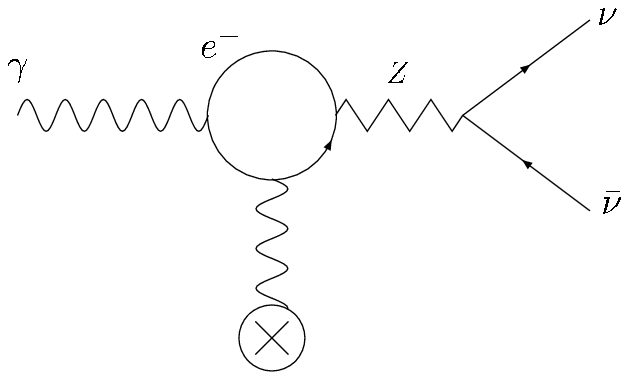}
\caption{The interaction of photons with a neutrino-antineutrino pair in a magnetic field proceeds through
two intermediate states: an electron (which has electric charge and weak charge) and and a
$Z$ boson (which has no electric charge).
If the photon energy is smaller than the electron and $Z$ mass, the amplitude of this process is suppressed by the
electron and the $Z$ masses.
Other possibilities for this transition involve $W^\pm$ bosons.}
\label{ggnu}
\end{figure}

The feebleness of the interactions between photons and neutrinos is, however, due to two reasons.
Firstly, it arises from the fact that, as their name indicates, neutrinos are neutral, i.e. they have no electric charge,
and therefore do not respond to the electromagnetic force.
Secondly, it originates from the above-mentioned necessity of involving very massive mediators.
Therefore neutrinos can interact with photons only through two intermediate steps, involving two ``mediators".
In Fig.~\ref{ggnu} we have depicted a diagram of one of the several possibilities.
There we see a photon interacting with an electron which can then interact with a $Z$ boson to
produce two neutrinos.
The mediation of such a number of so-called virtual particles makes the conversion of a photon into a neutrino pair even less likely than into a single graviton, cf.~\cite{Ahlers:2008jt,Gies:2009wx}.

Let us recapitulate what we have learnt until now.
Photons can shine through a wall (in a tricky sense) if they can convert into very weakly interacting
particles before and be regenerated after.
Within the standard model of particles physics, the graviton and a neutrino-antineutrino pair are the best suited candidates
for the intermediate step.
In order to perform the conversion before and after the wall we need the interplay of a ``mixing agent" that matches the
quantum numbers. We have pointed out that magnetic fields can be used as mixing agents to provide missing angular momentum.
We have shown that the photon-graviton and photon-neutrino pair conversion is extremely inefficient.
In the case of the graviton, we have not explained why (we still do not have a successful explanation of why gravity is so weak)
but in the case of the neutrinos we have seen that their interaction with photons are very suppressed due to the
necessity of
very massive mediators.
Therefore, if that were all, we would be forced to conclude that there are good chances to never observe
these effects in a laboratory.

Fortunately it seems that this could be not the case.
In the last decades we have become more and more convinced that the SM cannot be the end of the story.
Despite its success, there are both theoretical and observational motivations to
believe that the SM describes just a small component of nature's complexity.

\subsection{A hymn for physics beyond the SM}

There are two kinds of arguments to believe that there is new physics (mainly meaning new particles) beyond the SM:
observational and theoretical.

We have very strong observational evidence from the very active fields of astrophysics and cosmology
that only $\sim 4\%$ of the universe's energy is constituted by identified types of matter (baryons and electrons in different forms).
Around 22\,\% of the energy budget is made of a yet unidentified type of matter, that we call ``dark matter''.
The experimental evidence comes from many sources, but so far we have only been able to pin down two
properties of this substance: it interacts gravitationally but not through electromagnetic or strong forces and,
if it is matter made of particles, they should be non-relativistic, i.e. they should have velocities much smaller than the speed of light.
If dark matter is made of more than one type of particles, these characteristics should apply to the dominant
contribution and not necessarily to the same extent to the rest.

Note that the requirement that the velocity is small does not necessarily imply that the mass is large:
dark matter could be made of a small number density of very massive particles or, alternatively, of a large number
density of very light particles. The prejudice against light particles as dark matter relies on the assumption
that they are thermally produced in the big bang, an attractive, but not necessarily true hypothesis.

Even more mysterious is the claim that yet another 74\,\% of the universe energy budget
is made of something for which we don't have a better name than ``dark energy''.
These two unexplained forms of energy hint for for physics beyond the SM.

Let us now turn to purely theoretical arguments. The standard model of particle physics suffers, for a theoretician's taste, from some `aesthetical' problems.
With only a few parameters -- like particle masses and their coupling strengths -- it describes very well the outcome of laboratory experiments so far, but it does not provide for an explanation of the values these parameters take.

Moreover, with the evolution of our theories of particle physics, scientists have certainly
developed a strong bias towards the concepts of unification and symmetry.
In the late 1800's, the Maxwell equations combined the electric and magnetic interactions in a single framework showing that they are two
components of a unified entity, the electromagnetic field tensor. A further step was taken
in the 1960's when Glashow, Salam and Weinberg constructed a theory
unifying the electromagnetic and weak nuclear forces.

The standard model is certainly not yet a very unified or symmetric theory.
Following the above lines of thought, particle physicists have proposed different frameworks for further
forms of unification. The so-called Grand Unified Theories can explain the electroweak and
strong forces as two components of a new, more symmetric, unified interaction.
There is also an absolute lack of symmetry in the SM when we realize that the force carriers (photons, $W^\pm,Z$ bosons and gluons) are bosons while the matter particles (leptons and quarks) are fermions.
A new symmetry called supersymmetry (SUSY) can be added to the SM which results in doubling the number of particles, such that for each boson in the SM there is an associated fermion and viceversa\footnote[5]{The fact that these new particles, called super-partners, have not been observed yet can be accommodated when supersymmetry is broken by giving them relatively large masses.}.
This apparent complication of the SM pays off very generously, at least in three ways.
First, SUSY is the most general space-time symmetry containing the Poincar\'e group
(translations and proper Lorentz transformations) and allowing a consistent 4-dimensional quantum field theory
(Haag-Lopuszanski-Sohnius theorem).
Second, SUSY decreases the importance of quantum corrections to the mass of the Higgs boson (the only particle in the SM that is not yet discovered, but is believed to be found at the Large Hadron Collider).
A very basic line of arguments suggests that these corrections are extremely important and tend to rise the Higgs boson mass to values where the SM in itself is not well defined. With SUSY, these corrections are cancelled.
Third, SUSY predicts a good dark matter candidate, the lightest SUSY partner, if a series of reasonable assumptions are made.

An even more annoying aspect of the standard model is that it does not provide a consistent
framework to calculate quantum gravity effects.
Within every-day life physics, these effects are completely negligible because gravity is extremely weak.
However, at very small distances, of the order of the so-called Planck length ($\sim 10^{-35}$~m !), gravitational
interactions become comparable to electroweak and strong forces, such that quantum gravity effects become relevant.
The uncertainly principle relates these small distance effects to an incredibly high energy scale known as the
Planck mass, $M_{\rm Pl}=1.2\times 10^{19}$~GeV.

Currently, the most promising candidate for extending the standard model to include quantum gravity is
string theory. In string theory, the concept of point-like particles is replaced by the concept of strings
propagating or ``living'' in nine spatial dimensions.
Having extra dimensions is a requirement of string theory
to be consistent.
Six dimensions out of nine are not observed, which could be interpreted as a drawback for string theory.
However there is the possibility that these extra dimensions are curled up and thus much smaller than our
standard three dimensional space.
For instance, a natural scale of the sizes of the extra dimensions would be the Planck length.

For macroscopic objects like us these dimensions appear unresolved in every-day life.
Since we cannot determine the position of strings in the extra-dimensions we can build an effective theory
where all the extra dimensions are shrunk into our three dimensional world.
This process is called compactification.

There are many ways of compactifying string theory (this research field is very lively) and the outcome
can be very different.
However, it seems to be a common feature of realistic compactifications that they always include
a large number of additional particles, corresponding for instance to low-lying excitations of the strings or the geometry of the extradimensions itself.
To close for the moment our little chapter about string theory, let us only remark that it includes very naturally the concepts of unification of the fundamental forces and requires in most cases the existence of supersymmetry.

It is interesting to remark that although the motivation for unifying and finding new symmetries in the dynamics of particles is to obtain a simpler picture of nature, this simplification is not completely obvious from all perspectives.
It is a common feature of all the ideas exposed above that they imply the existence of a large number of still undiscovered particles.
These particles give rise to dynamics at characteristic lengths that we have not been able yet to explore, or equivalently to energy scales that we haven't achieved in laboratories (essentially above the 100 GeV ballpark).
SUSY partners, for instance, would have masses of the order of TeV and particles related to grand unification are related to energy scales around $10^{15}$~GeV. As already mentioned, quantum gravity points to a dynamical scale much larger, the Planck mass.
The simplification of introducing more particles only becomes evident when the energies involved are larger than the above-mentioned scales.

The unexplored territory in energy scales is so huge (maybe from 100 GeV to $M_{\rm Pl}$) and the possibilities so many that we cannot avoid the thought of having too few pieces of the puzzle to discard those completely unexpected.
When one combines this idea with the enormous difficulties that we have to face to explore this high energy frontier, an unavoidable question arises: \\

\emph{Can there be low energy dynamics related with physics at these unexplored high energy frontiers?}\\

The answer is a sound yes. The reader may remember the two cases mentioned in the introduction in which the coupling of SM particles (photons) with another low mass particle can produce our beloved light-shining through walls effect. Gravitons and neutrinos interact so weakly because their only connection with SM particles happens at very small length scales. The weakness of gravity, for which so far we don't give an additional explanation but it is most likely related to dynamics at the Planck length, did the job in the first place while the mediation of the massive electroweak bosons $W$ and $Z$ took the responsibility in the second.

Nothing in this ambitious theoretical structure sketched above prohibits that new light particles beyond the SM exist. The only restriction, which comes from ordinary observations and day-life phenomena,  is that they should not have SM  charges (strong or electroweak), i.e. they have to populate what we call a ``hidden sector''.
Actually these ``hidden sectors'' arise quite naturally in string theory and are required to explain the unobservation of SUSY partners (technically speaking to break SUSY at low energies).
Low mass hidden particles are therefore theoretically well motivated WISPs (i.e. extremely weakly interacting slim particles) candidates that could in principle allow for shining light through walls.

The outline of this review is as follows. In section~\ref{WISPs} we describe what kind of WISPs
are predicted in popular extensions of the standard model and discuss their couplings to
photons, required to allow shining light through walls. In section~\ref{sec:oscillations}
we review the physics of the photon-WISP conversions (oscillations).
In section~\ref{sec:exper}, we get our hands dirty and go to the lab, to describe the past and present
experimental efforts to detect light shining through walls.
We also describe, in section~\ref{sec:astrocosm},  how light could shine across other physical boundaries leading to interesting
effects in astrophysics and cosmology. In fact, there are various observations which seem to point to WISPs with
coupling strengths which may be probed in the next generation of light shining through a wall experiments, as will
be summarized in section~\ref{sec:next}.
Finally, in section~\ref{sec:conclusions}, we discuss the implications of a future discovery and present our conclusions.

\newpage
\section{The WISP zoo}\label{WISPs}
%
%

There is a strong prejudice among theoretical physicists against very small masses
because quantum corrections tend to erase big mass hierarchies.
However, the standard model itself is much lighter than the GUT or Planck scales so
it provides us with a very nice example that nature can create very nice exceptions to the
above rule.

One can control this destabilisation property of quantum corrections with appropriate symmetries.
Turning the argument around, we can organise the possible WISPs to appear in the hidden sector by
listing the symmetries that we know can protect particle masses.
In this section we briefly review the most famous candidates that fit the above-mentioned program.
There are, however, other possibilities (remember... expect the unexpected!) that we shall introduce as
a coda for our WISP compilation.

Besides a brief theoretical motivation we take the opportunity to write down the
Lagrangians that describe the WISP interactions with photons. Indeed, we shall restrict ourselves to
the type of interactions which is most relevant for shining-light-through-walls: photon-WISP mixing.
In the next section we will motivate this choice and show how to compute the probability of LSW in each case.

\subsection{Axions and axion-like particles}

As the first and paradigmatic example we find the axion and other axion-like particles (ALPs),
the smallness of their
mass being related to a shift symmetry of the theory under the replacement of the corresponding field
by an additive constant, $\phi(x)\to \phi(x) + {\rm const}$.
Such a symmetry forbids explicit mass terms, $\propto m_\phi^2 \phi^2$,
in the Lagrangian, rendering the particles corresponding to the excitations of the field $\phi(x)$  massless.
Moreover, it leads to the fact that the couplings of axions and ALPs to
standard model particles can only occur via derivative couplings, $\propto \partial_\mu \phi/f_\phi$, leading to a strong
suppression of their interactions at energy scales below $f_\phi$, the so-called $\phi$ decay constant.
ALPs are represented by scalar fields and therefore lead to spin-zero particles.

ALPs are not always exactly massless: often
there are some terms in the low energy effective Lagrangian which break the shift symmetry explicitly. If a term of order $\Lambda^4$ ($\Lambda$ is then an energy scale related to the dynamics that do not respect the shift symmetry) appears in the axion potential, then the ALP mass turns out to be non-zero, but still parametrically small if $\Lambda\ll f_\phi$,
\begin{equation}
m_\phi\sim
\frac{\Lambda^2}{f_\phi}\ll f_\phi.
\end{equation}
Note that a high energy scale $f_\phi$ suppresses both the axion interactions and its mass,  therefore ensuring the WISPy nature of ALPs.

Let us now turn to the case of the proper axion $a(x)$, sometimes called Peccei-Quinn axion or
QCD axion~\cite{Peccei:1977hh,Weinberg:1977ma,Wilczek:1977pj}. In this case,  the shift symmetry is broken at the quantum level by the colour anomaly, producing a term in the Lagrangian
\begin{equation}
{\mathcal L} \supset
\frac{\alpha_s}{4\pi}\, \frac{a}{f_a}\, {\rm tr}\, G_{\mu\nu} {\tilde G}^{\mu\nu} \equiv
\frac{\alpha_s}{4\pi}\, \frac{a}{f_a}\,
\frac{1}{2}\,\epsilon^{\mu\nu\alpha\beta}\,{\rm tr}\, G_{\mu\nu} G_{\alpha\beta},
\label{topterm}
\end{equation}
where $\alpha_s$ is the strong coupling and $G$ is the gluonic field strength.
This interaction creates a non-perturbative potential for the axion field $a(x)$ basically because of the strong
self-interactions of gluons. Most importantly, the minimum of the potential
cancels the so called $\theta$-term in the Lagrangian of quantum chromodynamics (QCD),
\begin{equation}
{\mathcal L}_{\rm CP-viol.}^{\rm QCD} =
\frac{\alpha_s}{4\pi}\, \theta\, {\rm tr}\, G_{\mu\nu} {\tilde G}^{\mu\nu} ,
\label{thetaterm}
\end{equation}
solving the so called strong CP puzzle\footnote[6]{The interested reader is encouraged to read~\cite{Peccei:2006as}, and/or the brief review of the Particle Data Group on axions and similar particles~\cite{Amsler:2008zzb},
for a comprehensive review on axions and the strong CP problem. For an even
more pedagogical and illustrative review see Ref.~\cite{Sikivie:1995pz}.}, namely why the value of $\theta$ inferred from neutron electric dipole moment experimental searches is so small ($\left|\theta\right|\lesssim 10^{-10}$) while theoretically there is no preference for any specific value in the range $(0,2\pi)$.
Note that the axion mechanism is valid regardless of the value scale $f_a$ (as long as $f_a\gg \Lambda_{\rm QCD}$). In particular, $f_a$ can be much larger than the electroweak scale~\cite{Kim:1979if,Shifman:1979if,Dine:1981rt, Zhitnitsky:1980tq}.

The expansion of the potential around this minimum provides the axion with a non-zero mass,
\begin{equation}
\label{axionmass}
m_a\sim \frac{f_\pi m_\pi}{f_a} \sim { {\rm meV}}
     \times
     \left(
     \frac{10^{10}\, {\rm GeV}}{f_a}\right),
\end{equation}
where $f_\pi$ and $m_\pi$ are the decay constant and mass
of the pion, respectively.

In addition to the derivative couplings to standard model matter and
most importantly in the context of light shining through a wall matters,
axions and ALPs can also have anomalous couplings to electromagnetic fields (photons),
in analogy to Eq.~(\ref{topterm}),
\begin{equation}
{\mathcal L} \supset
\frac{g}{4}\, \phi\, F_{\mu\nu} {\tilde F}^{\mu\nu} =- g\, \phi \, \vec E \cdot \vec B
,
\label{twogammacoupl}
\end{equation}
where $F$ is the electromagnetic field strength and $\vec E$ and $\vec B$ are the electric and magnetic fields.
The effective coupling is expected to be of order
\begin{equation}
\label{axioncoupling}
g\sim  \frac{\alpha}{2\pi f_\phi }\sim 10^{-12}\ {\rm GeV}^{-1}          \left(
     \frac{10^{9}\, {\rm GeV}}{f_\phi }\right),
\end{equation}
where $\alpha$  is the fine-structure constant.

Axions and ALPs arise in many extensions of the standard model as the number of particles is large and this somehow tends to increase the number of possible shift symmetries.
Most importantly, axions and ALPs are \emph{generic} in string theory with $f_\phi$ being of the order of the string scale $M_s$ ($M_s$ is just the inverse of the fundamental string length)~\cite{Witten:1984dg,Conlon:2006tq,Svrcek:2006yi}.
In fact, when compactifying the six extra spatial dimensions of string theory they arise quite naturally as
nearly massless excitations of certain types of strings.
Moreover,
couplings in the low-energy effective Lagrangian like those in Eqs.~(\ref{topterm}) and (\ref{twogammacoupl}),
are also unambiguously obtained from the scattering amplitudes of strings at low energies.
Thus, string compactifications suggest plenty of candidates for axions and axion-like
WISPs. Typical values of $M_s$, varying between $10^{9}$ and $10^{17}$~GeV, for intermediate scale and GUT scale strings, respectively, suggest values of $f_\phi$ in the same range.

In a background magnetic field\footnote[7]{The same holds in an electric field but this case is less interesting from a practical point of view.}, $\vec B_{\rm ext}$ the two photon coupling in Eq.~(\ref{twogammacoupl}) behaves as a photon-ALP non-diagonal mass, a so called \emph{mass mixing term}. This mixing term leads to the phenomenon of photon-ALP oscillations, to be described in Sec.~\ref{sec:oscillations}, in a very similar fashion to neutrino flavour oscillations (see the corresponding section in
Ref.~\cite{Amsler:2008zzb} for a pedagogical review).
In fact, using a plane wave propagating in the $z$-direction for the photon field $A^\mu=A_0^\mu e^{i \omega (t -z)}$ and a magnetic field in the $x-y$ plane one obtains
\be
\label{ALPmixing}
{\cal L}_{\rm mixing} = -i g \omega   (\vec B_{\rm ext} \cdot \vec A    ) \phi ,
\ee
where we have chosen the radiation gauge $\vec \nabla \cdot \vec A=0$. We see that this mixing term only holds for the photon polarization aligned with the external magnetic field.

The phenomenology arising from the coupling in Eq.~(\ref{twogammacoupl}) stays very much unchanged also for for a second type of two photon coupling,
\begin{equation}
{\mathcal L} \supset
\frac{g}{4}\, \phi\, F_{\mu\nu} {F}^{\mu\nu} = g\, \phi\,\( {\vec B}^2-{\vec E}^2\) ,
\label{twogammacoupl2}
\end{equation}
which in a magnetic field produces mixing between the axion-like particle $\phi$ and the photon component which is \emph{perpendicular} to the external field.
Particles featuring this coupling could be for instance quintessence fields~\cite{Wetterich:1987fm,Ratra:1987rm,Frieman:1995pm,Khoury:2003aq} or particles that arise from excitations of fields governing the sizes of extra dimensions (moduli) or gauge couplings (dilatons) in string theories.
Such particles are in principle subject to strong constraints from deviations of Newton's law~\cite{Dupays:2006dp}. This topic is in itself vast and a proper account of the different experiments and theoretical solutions is beyond the scope of this review. Let us just note that there are some models, i.e. some specific types of axion-like-particles that overcome these problems, cf. for instance~\cite{Masso:2006gc,Jaeckel:2006xm,Brax:2007ak}.

The couplings Eq.~(\ref{twogammacoupl}) and (\ref{twogammacoupl2}) respect CP (combination of charge conjugation and parity) symmetry if the ALP $\phi$ is pseudoscalar in the first case ($\phi(-x)=-\phi(x)$) or scalar ($\phi(-x)=\phi(x)$) in the second\footnote[8]{A brief explanation of this can be found in~\cite{Redondo:2008tq}.}.
Axions should be pseudoscalar if they are to solve the strong CP problem, while moduli and dilatons are pure scalar fields.
However, in principle ALPs can arise in sectors where CP is not respected and can therefore have a mixture of the two couplings, Eq.~(\ref{twogammacoupl}) and (\ref{twogammacoupl2}), see for instance~\cite{Hill:1988bu} where these fields with undefined CP properties are called \emph{schizons}.

\subsection{Hidden-sector photons}

A second interesting example of inhabitants of the WISPs zoo are hidden-sector photons (HPs), sometimes
also called hidden photons, paraphotons~\cite{Okun:1982xi}, or  dark photons. Similar to our familiar photons -- the
force carriers of electromagnetic interactions -- which are gauge bosons of the electromagnetic
gauge group U(1), HPs
are gauge bosons of an extra local Abelian U(1) gauge symmetry in the hidden sector.

Analogously to the global shift symmetry, which forbids a mass term for ALPs, the
gauge invariance usually forbids a mass term for gauge fields. A non-zero mass may however
be induced by spontaneous breaking of the symmetry via the Higgs mechanism.
In this case, in addition to the massive HP, also a hidden Higgs particle would be predicted.
This is just like in the standard model where a complex Higgs SU(2) doublet is used to give mass to the three weak bosons and one extra field remains, the Higgs particle.
In contrast to the Higgs in the standard model, which breaks a non-Abelian gauge symmetry (SU(2)$\times$U(1)), in the hidden U(1) case the mass of the corresponding Higgs particle can be arbitrarily high, even can be formally taken to infinity, realising the so called St\"uckelberg limit.
In this case the hidden Higgs disappears from the low energy effective field theory.
Not only this is quite a generic case from the theoretical perspective, but it also turns out to be phenomenologically the most interesting case when we consider low mass HPs. Indeed, the presence of a low mass hidden Higgs is subject to very strong constraints~\cite{Ahlers:2008qc} which render the observation of light shining through walls (LSW) in near future experiments quite unlikely.

Extra U(1) gauge factors are ubiquitous in well motivated extensions of the SM, most notably in string compactifications.
Even if the paradigm of a highly symmetric high energy theory is just to unify particles into representations of a large-rank local gauge group, such as SU(5), SO(10), SO(32) or ${\rm E}_8\times {\rm E}_8$, the phenomenological fact that at low energies these large
gauge symmetries are broken leaves us with potentially many lower rank gauge symmetries. Interestingly, U(1)'s are the lowest-rank local symmetries so therefore potentially the most numerous, and of course, some of them can be hidden.

For example, in the standard compactification of the ${\rm E}_8\times {\rm E}_8$ supergravity theory
based on the heterotic string
from 9 to 3 spatial dimensions,
the standard model SU(3)$\times$SU(2)$\times$U(1) gauge group is embedded in the first ${\rm E}_8$ factor,
whereas the second ${\rm E}_8$ factor comprises a ``hidden gauge group", which interacts with
the first ${\rm E}_8$ factor only gravitationally. This second ${\rm E}_8$ factor may be broken in the
course of compactification to products of non-Abelian SU(N) and Abelian U(1) gauge groups.

In summary, light hidden-sector U(1)s are indeed very well motivated WISP candidates. Their dominant interaction with
the particles in the visible sector occurs via kinetic mixing with the photon,
encoded in the following term in the low-energy effective Lagrangian,
\begin{equation}
\mathcal{L} \supset
-\frac{\chi}{2} F_{\mu \nu} X^{\mu \nu}
,
\label{LagKM}
\end{equation}
with $X_{\mu\nu}$ denoting the hidden U(1) field strength.
In fact, kinetic mixing is expected to be small because the mixing parameter $\chi$ is generated at one-loop,
$\chi \lesssim e g_h/(16 \pi^2)\lesssim 10^{-3}$,
by the exchange of heavy messengers that couple both to the electromagnetic U(1), with a strength corresponding to
the unit of electric charge $e$, as well as to the hidden U(1), with a strength $g_h$~\cite{Holdom:1985ag}.
One should note that $10^{-3}$ is a typical number but there are many models that predict much smaller
values. If we consider string theory~\cite{Dienes:1996zr,Abel:2006qt,Abel:2008ai,Goodsell:2009xc,Goodsell:2010ie}, we find that when considering all the possible messengers that can generate kinetic mixing there can be cancellations between the contributions of them, suppressing $\chi$ sometimes down to the level of $10^{-17}$.
In models where the U(1) field arises from massless excitations of branes which wrap large cycles in the extra dimensions, the hidden gauge coupling itself can be enormously suppressed reducing the kinetic mixing down to the $10^{-14}$ level even if there are no cancellations between the messengers.

The kinetic mixing term, together with a non-zero HP mass, leads to photon-HP oscillations, as we will
show now.
We start with the low energy Lagrangian for the simplest U(1) extension of the SM, namely
\bea
{\cal L}  \supset -\frac{1}{4}F_{\mu\nu}F^{\mu\nu}-\frac{1}{4}X_{\mu\nu}X^{\mu\nu} -\frac{\chi}{2}X_{\mu\nu}F^{\mu\nu}+\frac{1}{2}\muu^2X_\mu X^\mu
+ ej_{\rm em}^\mu A_\mu +g_h j_{\rm hid}^\mu X_\mu,
\eea
where $A_\mu, X_\mu$ are the electromagnetic and hidden vector potentials, such that $F_{\mu\nu}=\partial_\mu A_\nu-\partial_\nu A_\mu$ and $X_{\mu\nu}=\partial_\mu X_\nu-\partial_\nu X_\mu$.
We have already written the HP mass, $\muu$, without specifying its source and the coupling of photons and HPs to their associated currents, the electromagnetic current $j_{\rm em}^\mu$ and a possible hidden current $j_{\rm h}^\mu$ composed of particles charged under the hidden U(1) (to be discussed in the next section).
The sequence of redefinitions
\be
X_\mu \to X_\mu - \chi A_\mu \quad ; \quad A_\mu \to \frac{1}{\sqrt{1-\chi^2}}A_\mu \quad ; \quad e\to e\sqrt{1-\chi^2}
\quad ; \quad \chi\to \chi\sqrt{1-\chi^2} \quad ;
\ee
removes the kinetic mixing term leading to
\bea
\label{HPlag}
{\cal L}  &\supset& -\frac{1}{4}F_{\mu\nu}F^{\mu\nu}-\frac{1}{4}X_{\mu\nu}X^{\mu\nu} +\frac{1}{2}\muu^2(X_\mu-\chi A_\mu) (X^\mu -\chi A_\mu)\\ \nonumber
&+& ej_{\rm em}^\mu A_\mu +g_h j_{\rm hid}^\mu (X_\mu-\chi A_\mu).
\eea
Note that in this representation there appears a mass mixing term,
\be
\label{HPmixing}
{\cal L}_{\rm mixing} = -\chi \muu^2 A_\mu X^\mu .
\ee
As mentioned in the last section, the mass mixing term is the key ingredient for having photon-WISP oscillations.
In this case, the mixing is proportional to the HP mass squared and the kinetic mixing and does not require a mixing agent like a magnetic field, required for photon-ALP oscillations.

\subsection{Minicharged particles}

Particles charged under the hidden U(1) appear as having a small interaction with photons,
${\cal L}\supset - g_h  \chi j_{\rm hid}^\mu A_\mu$ (recall the analogy with the $A_\mu$ coupling to the electromagnetic current in Eq. (\ref{HPlag})).
Hence, the hidden particle appears to have an electric charge of value
\be
Q = \frac{g_h \chi }{e},
\ee
which can be very small if either $\chi$ or $g_h$ (or both) are small.
This possibility was originally discussed by Holdom in Ref.~\cite{Holdom:1985ag} where those particles
were dubbed milli-charged particles since it was taken for granted that the natural value of
$g_h$ was to be similar to the electric charge and the kinetic mixing is naturally of order $10^{-3}$.
However, as highlighted before, the kinetic mixing can be much smaller and, therefore, these particles have been
re-baptized as mini-charged particles (MCPs).

These particles can be very light for a number of reasons.
For instance, mass terms can be forbidden by chiral symmetries (just as it happens in the SM) or protected by low energy supersymmetry in the hidden sector.

The existence of light particles of small (unquantised) electric charge is however not tied to the existence of hidden photons and the kinetic mixing mechanism outlined above. In models of extra dimensions, the smallness of the charge could simply arise because of spatial separation of the SM particles and the hidden sector~\cite{Batell:2005wa}.

The existence of MCPs can lead to light-shining-through walls in at least two ways.
In the minimal scenario, the photons shone against the wall can convert into a virtual MCP particle antiparticle pair that can make it through the wall and coalesce after to form a photon. This possibility, similar to the neutrino intermediate state of Fig.~\ref{LSWsm}, was studied in \cite{Gies:2009wx} where it was found to be extremely small. For instance, for MCPs of very small mass, and a wall thickness of the order of the MCP Compton wavelength, $d \sim 1/m_{\rm MCP}$,
\begin{equation}
P_{\gamma\to\gamma}(\omega\gg 2m_{\rm MCP}) \simeq
\frac{\alpha^2 Q^4}{9 \pi^2} \, \ln^2 \frac{\omega}{2 m_{\rm MCP}}\ ,  \label{eq:5}
\end{equation}
where $\omega$ is the photon energy.

\begin{figure}[t!]
\begin{center}
\includegraphics[width=14cm]{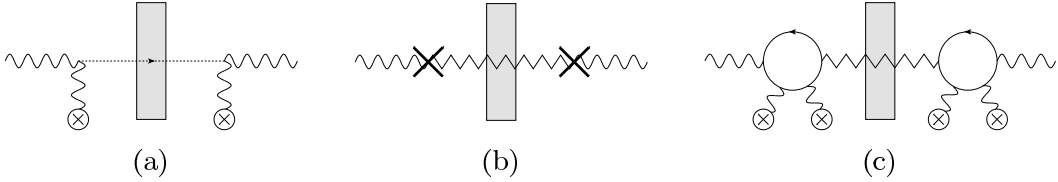}
\caption{Explicit processes contributing to LSW for various WISPs.
From left to right we have photon -- ALP, photon -- hidden photon
and photon -- hidden photon oscillations facilitated by
MCPs.}\label{fig:LSWcand}
\end{center}
\end{figure}

There is however another interesting way in which MCPs can lead to LSW.
In the hidden photon model of the previous section the MCPs can mediate a transition between a photon and a hidden photon in a loop,  the HP making it through the wall, see Fig.~\ref{fig:LSWcand} (c).
If the HP mass is very small (or simply zero) this contribution will dominate the LSW probability.

Particularly interesting is the case when there is a magnetic field present in the conversion region.
The minicharged particles get non-trivial propagation properties inside the magnetic field because
of their small charge. This makes the mixing of photons and hidden photons itself inheriting new properties: it becomes polarization dependent (the polarizations parallel and perpendicular to the magnetic field mix differently) and its strength is also enhanced.
The loop diagram giving rise to the mixing can be computed for MCPs of arbitrary small charge and mass, under a number of approximations~\cite{Gies:2006ca,Ahlers:2006iz,Ahlers:2007rd}. Its effects can be described by an effective polarization tensor
$\Pi_{\mu\nu}$ in the Lagrangian which is equivalent to a non-diagonal index of refraction (a non-diagonal mass term again). Integrating out the MCP fields we would obtain
\bea
\label{LMCP2}
{\cal L} &\supset & (X^\mu-\chi A^\mu) \Pi_{\mu\nu}(X^\nu-\chi A^\nu) \\ \nonumber
&\equiv &
-\frac{1}{2} m^2_{A,||}A_{||}^2 -\frac{1}{2} m^2_{X,||}A_{||}^2
-\frac{1}{2} m^2_{A,\perp}A_\perp^2 -\frac{1}{2} m^2_{X,\perp}A_\perp^2
-\frac{1}{2}\delta m^2_{||}A_{||}X_{||} -\frac{1}{2}\delta m^2_\perp A_\perp X_\perp ,
\eea
where $||,\perp$ labels stand for the photon and HP polarizations aligned or perpendicular to the magnetic field.
From Eq.~(\ref{LMCP2}) we find that the masses and mass mixings satisfy
\be
m^2_X::\delta m^2:: m^2_A = 1 ::\chi :: \chi^2 ,
\ee
so they are all given in terms of $m^2_{X,||}$ and $m^2_{X,||}$.
The formulas of $m^2_{X}$ as function of the magnetic field and frequency are quite involved so we direct the interested reader's curiosity to the references~\cite{Gies:2006ca,Ahlers:2006iz,Ahlers:2007rd} for a proper satisfaction.
Fortunately, there is a limit (maybe the most interesting case) when the expressions are tractable, namely when the MCP mass is sufficiently small such that its propagation in the magnetic field becomes non-perturbative. In this case we find the mass mixing terms
\bea
m^2_{X,||} &=& 	3\frac{3^{2/3}\sqrt{\pi}}{14\ 2^{1/3}}\frac{\Gamma\(\frac{2}{3}\)^{2}}{\Gamma\(\frac{1}{6}\)}	\(\omega e Q B\)^{2/3}	 +i \frac{1}{12 \Gamma\(\frac{1}{6}\)\Gamma\(\frac{13}{6}\)}\(\omega e Q B\)^{1/3}		,			\\
m^2_{X,\perp} &=& \frac{2}{3}m^2_{X,||} ,
\eea
where $\omega$ is the HP frequency, $e$ the electron charge, $B$ the magnetic field strength and we have considered the MCP particle to be a Dirac spinor.

Note that the mixing is complex and therefore it leads to HP (and photon, to a lesser extent) disappearance just as an imaginary index of refraction parameterizes photon absorption.
This is due to the fact that HPs can decay into an MCP pair while propagating in the magnetic field.
Interestingly, this can happen even if the HPs were massless to start with, because its dispersion relation and that of MCPs has been altered by the magnetic field.
A very similar phenomenon exists in the context of QED. By means of it, a photon propagating in a strong enough electric field can ``decay'' into an electron/positron pair. This process was measured (in a tricky sense) in the SLAC E-144 experiment~\cite{Burke:1997ew} in the perturbative regime but the non-perturbative production has escaped experimentalists so far and as the majority of predictions of non-perturbative QED remains unexplored and untested. In fact, the existence of very light MCPs and their discovery could teach us much about the ``known'' theory of QED
in the up to now untested range of ultra-strong fields~\cite{Gies:2006hv,Gies:2008wv}.

\subsection{Other WISPs}

The imagination of particle physicists is not exhausted by axions, ALPs, hidden photons and
minicharged particles as WISP candidates. It would be beyond the scope of this review to present
a complete list of all of them. Here, we just mention three more and explain, why, in the context
of LSW, they are less discussed in the literature.

The {\em gravitino} -- the spin 3/2 super-partner of the spin 2 graviton -- is definitely a well motivated WISP candidate in the context of supergravity -- the supersymmetric extension of the
standard model including gravity. Its mass is given in terms of a see-saw relation between the SUSY breaking scale
$\Lambda_{\rm SUSY}$ and the Planck scale,
\be
m_{3/2} = \frac{\Lambda_{\rm SUSY}^2}{M_{\rm Pl}}.
\ee
Clearly, for $\Lambda_{\rm SUSY}$ below 100~TeV, the gravitino mass is in the sub-eV range.
Moreover, its interactions with photons are suppressed by inverse powers of the SUSY breaking
scale~\cite{Brignole:1996fn,Luty:1998np,Clark:1997aa}.
However, for phenomenologically acceptable values of the latter ($\Lambda_{\rm SUSY}\gtrsim 1$~TeV),
eventual LSW effects from virtual gravitino pairs appear to be subdominant to the already tiny effects
from virtual neutrino pairs.

{\em Massive spin-2 particles} have also been discussed in the context of photon-WISP mixing~\cite{Deffayet:2000pr,Biggio:2006im}.
The best motivated ones arise in models with extra dimensions, as Kaluza-Klein (KK) modes of the graviton,
the small mass then associated with a possibly large size of the latter~\cite{ArkaniHamed:1998rs,Antoniadis:1998ig}.
Taking into account, however, the smallness of the effective photon-KK graviton mixing,
which is suppressed by the Planck mass, $\sim 1/M_{\rm Pl}$,
and the strong phenomenological lower bounds on the size of the extra dimensions, it appears that
the discovery potential of LSW experiments for KK gravitons is slim.

In scalar-tensor theories of gravity so-called {\em chameleons} -- scalar particles whose mass increases with
the local matter/energy density -- appear~\cite{Khoury:2003aq,Khoury:2003rn,Brax:2004qh}.
For chameleons, LSW does not work: the high density in
the wall increases the mass of the produced WISP and therefore creates a high potential barrier on which
the particle is reflected~\cite{Jaeckel:2006xm}.
However, by a slight change of the setup of LSW experiments, one may search
for chameleons which couple to photons by searching for an afterglow from chameleon-photon conversions
taking place before the wall after switching off the light source~\cite{Ahlers:2007st,Gies:2007su,Brax:2010jk},
see Refs.~\cite{Chou:2008gr,Upadhye:2009iv,Rybka:2010ah} for pioneering experimental results from such a setup\footnote[9]{Chameleon theories become strongly interacting inside dense environments and therefore less tractable. There are still nowadays serious problems of interpretation of chameleon afterglow experiments.}.

\section{Photon $\leftrightarrow$ WISP oscillations}\label{sec:oscillations}

In this section, we want to briefly explain the physics of photon-WISP oscillations.
This is the main mechanism on which LSW experiments rely for the production of WISPs
before the wall and the WISP reconversion into a photon after it.
As we shall see, photon-WISP oscillations present some advantages over other possible WISP production mechanisms.
The physical reason is that photon-WISP oscillations are processes that can occur \emph{coherently} over macroscopic distances.

\begin{figure}[tbp] 
  \centering{
  \includegraphics[width=16cm]{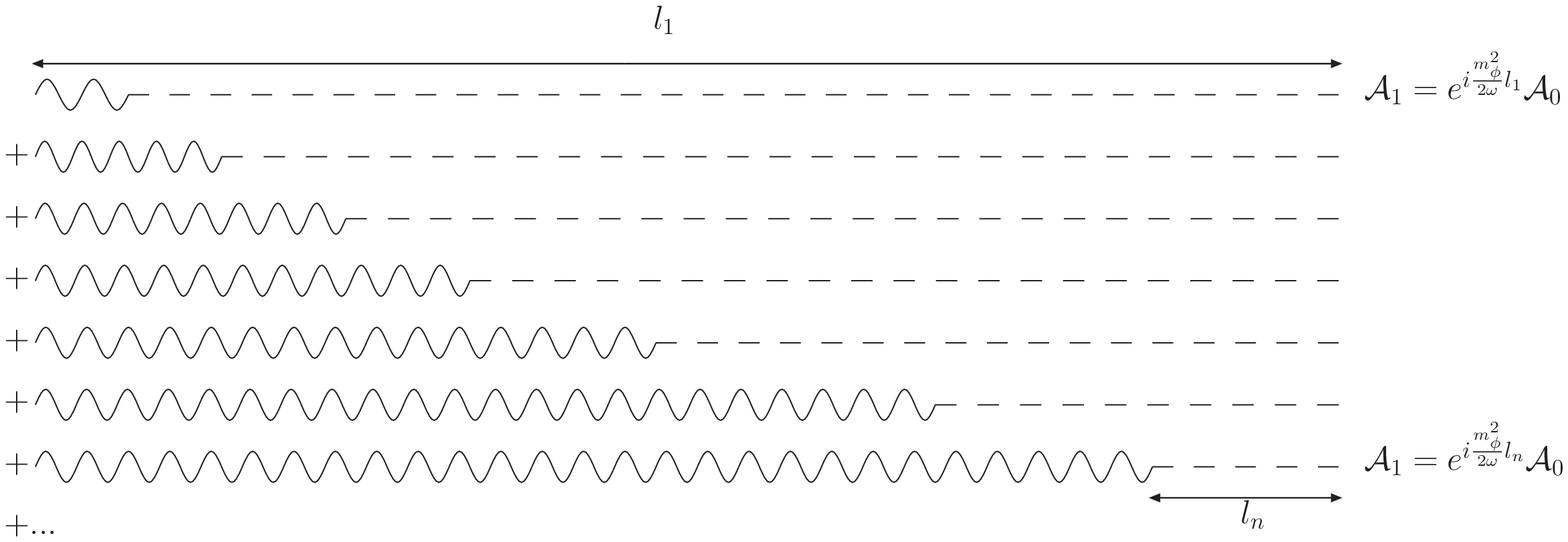}}
  \caption{The conversion of a photon into a WISP through mass mixing can happen at any position along the distance of the photon emitter and WISP receiver. The amplitude of all these processes differs only by a phase which depends on the relative ``speed'' of propagation of the photon and WISP waves. If this difference in phase velocities is small, all the amplitudes interfere constructively and enhance the photon-WISP probability.
This is what we call a coherent production mechanism. }
  \label{fig:coherent}
\end{figure}

Let us offer an appetizer to highlight this point.
We consider the conversion of a photon into an ALP in a constant magnetic field $B$ of length $L$.
In the cases of interest for this review these conversion processes are essentially 1-dimensional, i.e. the WISPs produced are collinear with the photons (or viceversa), so we will stick from now on to this case.
The probability amplitude of the photon WISP conversion (per unit length) can be read off directly from the mixing Lagrangian in Eq.~(\ref{ALPmixing}) as
\be
{\cal A}_0= i g B/2.
\ee
The key point is to realise that this conversion can happen at every position along the length $L$, but that conversions at different positions will have a phase difference because they travelled different distances as an ALP, see Fig.~\ref{fig:coherent}.
Using that the ALP wavenumber is $k_\phi=(\omega^2-m_\phi^2)^{1/2}$ or, in the relavistic limit,
$k_\phi\approx \omega-m_\phi^2/(2\omega)$, this phase difference is given by $m_\phi^2 l/(2\omega)$, where $l$ is the distance
traveled by the ALP.
Summing over the conversions at different values of $l$, which produce indistinguishable ALP final states, can be done by performing the integral
\be
\label{eq:ALPoscillationintegral}
{\cal A}(\gamma\to\phi )=  i\int_0^L
\frac{g B_\mathrm{ext}}{2} e^{i\frac{m_\phi^2}{2\omega}l} dl =
i\frac{g B_\mathrm{ext}\omega}{m_\phi^2}\left(1-e^{i\frac{m_\phi^2}{2\omega}L}\right) ,
\ee
which gives a conversion probability
\be
\label{Palp}
P(\gamma \to \phi )=|{\cal A}|^2 = 4\frac{g^2B_\mathrm{ext}^2 \omega^2}{m_\phi^4} \sin^2 \left(\frac{m_\phi^2 L}{4\omega}\right) .
\ee
Clearly, when $L<4\pi\omega/m_\phi^2$ the conversion probability is proportional to $L^2$, as expected from a coherent process along the whole length. Note that the interference can also be destructive when $m_\phi^2L/(4\omega)>\pi/2$ and can drive to zero the amplitude when this phase is an integer multiple of $\pi$.

We are now in a position to compare quantitatively the advantage of the coherent production of ALPs
with an incoherent production mechanism which we take as a beam dump.
Just as photons can convert into axions in an external magnetic field, they can do the same in the electric field of protons or electrons of a photon beam dump, via the so called Primakoff process.
Interference between the axion production off different target particles of the dump depends on the correlation between the different particle positions and cancels out for a completely uncorrelated ensemble (an ideal gas, for instance).
Moreover, there is also negative interference between positively and negatively charged particles because the amplitude is proportional to the scatterer's charge leading to the presence of screening\footnote[10]{The interested reader will find section 6.4 of Ref.~\cite{Raffelt:1996wa} very illuminating.}.
Being interested in a rough comparison with the oscillation formula, let us neglect all these effects\footnote[11]{
In highly ordered solids, one can in principle benefit from the coherent production in the different atoms in the crystal, like in Bragg scattering~\cite{Buchmuller:1989rb,Paschos:1993yf}}.  Then, the probability of conversion can be estimated as
\be
P(\gamma\to\phi )= \sigma_{\rm P} n_{\rm t} \lambda_{\rm abs} \approx  g^2 n_{\rm t} \lambda_{\rm abs} ,
\ee
where $\sigma_{\rm P}$ is the Primakoff cross section, $n_{\rm t}$ is the density of targets (nuclei and electrons) and $\lambda_{\rm abs}$ is the absorption length in the target.
Using a typical value $n_{\rm t}\sim N_A$ cm$^{-3}$, where $N_A\sim 10^{23}$ is Avogadro's number, we can compare with the probability in Eq.~(\ref{Palp}) in the coherent regime,
\be
\frac{P({\rm beam\ dump})}{P({\rm B\ field})}\approx \frac{n_{\rm t} \lambda_{\rm abs}}{B^2 L^2}
\sim 10^{-6} \(\frac{\lambda_{\rm abs}}{\rm mm}\) \(\frac{\rm T}{B}\)^2\(\frac{\rm m}{L}\)^2 .
\ee
Thus, the flux of axions from a beam dump is typically much smaller than the coherent production in a 1 meter long 1 Tesla magnetic field\footnote[12]{It is interesting to note that other electromagnetic fields made in laboratories can compete in $B\times L$ with our example here, for instance the fields created in ultrashort pulses of the most intense lasers.
A recent study highlighted this possibility~\cite{Dobrich:2010hi} finding that experiments of LSW type are not optimal 
 in this context while some other signatures of WISPs can be searched for.} (which is not the strongest and longest available magnet).

A second benefit from the coherent production is that magnetic fields have larger transversal extents than atoms and so the produced axions suffer less diffraction and the resulting axion beams are more collimated, which will facilitate their detection.

Finally, there is a third and most important reason not to dump the photons. If instead of a dump
we use a mirror we can recover these photons and redirect them again towards the wall, i.e. we can
recycle the photons with the use of two facing mirrors (what we call an optical cavity). This way a photon can be used many times (up to $10^5$ times!), enhancing the probability of making it through the wall in one of these attempts.
The use of optical cavities will be discussed later but we want to emphasize already here that it is an extremely advantageous technique which we cannot avoid to pursue in order to reach the ultimate sensitivity of the LSW technique.

Once convinced that the coherent production is a worthwhile production mechanism of WISPs let us move to consider a further generalization of the photon-WISP conversion probability.
If the transitions do not happen in vacuum but in a medium, the diffractive and refractive properties of light have to be taken into account. We can parametrize this by introducing an effective photon mass $m_\gamma$ and an absorption coefficient $\Gamma_\gamma$. They are related to the more commonly used complex index of refraction $n$ by
\be
\label{index}
2\omega^2(1-n) = m_\gamma^2 - i \omega \Gamma_\gamma .
\ee
Furthermore, we can include a WISP decay rate, $\Gamma_w$, for completeness.
The amplitude of $\gamma\to$WISP conversion is then, in full generality,
\bea
\nonumber
{\cal A}(\gamma\to {\rm WISP} ) &=&
\int_0^L i{\cal A}_0^w e^{i\frac{m_w^2-m_\gamma^2}{2\omega}l} e^{-\frac{\Gamma\gamma}{2}(L-l)}e^{-\frac{\Gamma_w}{2}l}dl \\
&=&\frac{2\omega {\cal A}_0^w}{m_w^2-m_\gamma^2-i \omega(\Gamma_\gamma-\Gamma_w)}
\left(e^{-\frac{\Gamma_\gamma}{2}L}-e^{-\frac{\Gamma_w}{2}L}e^{i\frac{m^2_w-m_\gamma^2}{2\omega}L}\right) ,
\eea
where $m_w$ is the mass of the WISP (ALP $\phi$, HP $\gamma^\prime$, \ldots),
so that the conversion probability is
\begin{eqnarray}
\label{generalprob}
\lefteqn{
P(\gamma\to {\rm WISP} )=}\\ \nonumber && \frac{(2\delta m^2)^2}{(m_w^2-m_\gamma^2)^2+\omega^2(\Gamma_\gamma-\Gamma_w)^2}
\left(e^{-\Gamma_\gamma L}+e^{-\Gamma_w L}-2
e^{-\frac{\Gamma_\gamma+\Gamma_w}{2} L} \cos\( \frac{(m^2_w-m_\gamma^2)L}{2\omega}\) \right) .
\end{eqnarray}
Here, $\delta m^2_w\equiv |2\omega  {\cal A}_0^w|$ denotes the WISP specific off-diagonal mass appearing in the Lagrangian.

Note that considering oscillations in a medium can be either extremely advantageous or harmful.
In vacuum, the amplitude of the oscillations reaches a maximum at $(2\delta m^2)^2/(m_w^4+(\omega\Gamma)^2)$ which is typically smaller than one. Actually we will see
that this is actually strictly the case when we consider a full version of the physics, because an additional factor $(2\delta m^2)^2$ appears in the denominator. Therefore, for $m_w^2\neq 0$, the amplitude of the oscillations is always $\leq 1$ and can be very suppressed if the WISP mass is large.

In a medium, the amplitude of the oscillations appears divided by the factor ${(m_w^2-m_\gamma^2)^2+\omega^2(\Gamma_\gamma-\Gamma_w)^2}$. If the medium is very dense, the corresponding photon effective mass
will be very large since normally $(n-1)$ is proportional to the density of particles in a medium.
Thus, for a given WISP mass, there is always a critical density for which the denominator starts to grow
as $m_\gamma^4+(\omega\Gamma_\gamma)^2$ and suppresses the amplitude of the oscillations, i.e.
the probability.

In order not to suppress photon-WISP oscillations for a given WISP mass $m_w$, the pressure of the ambient gas (considered to be ideal and with an index of refraction $n$ at standard conditions of temperature and pressure) has to be
\be
\(\frac{P}{{\rm mbar}}\) \ll 0.52 \(\frac{T}{273.15\,{\rm K}}\) \(\frac{{\rm eV}}{\omega}\)^2\(\frac{10^{-5}}{|n-1|}\)
\(\frac{m_w}{{\rm meV}}\)^2.
\ee
Therefore, if either the photon mass or the WISP mass is much larger than the mass mixing, the
probability is suppressed by the forth power of this ratio, which can be a very small number.

Fortunately, one can in principle revert the situation if one is able to match the photon and WISP masses and decay lengths. In this case we have perfect coherence of the WISP production along the length $L$.
Taking the limit $m_\gamma^2\to m_w^2$ and $\Gamma_\gamma=\Gamma_w$ one finds a very simple and expected formula for small $\delta m^2 L/\omega$,
\be
P(\gamma\to {\rm WISP} ) = \(\frac{\delta m^2L}{\omega}\)^2e^{-\Gamma_\gamma L},
\ee
which can be much larger than the $m_\gamma^2=0$ case.
Unfortunately, our formula is not well behaved in this case (the probability eventually blows up) and we will need a slight modification which we will comment later.
However, this does not spoil the conclusion that the case $m_\gamma^2=m_w^2$ is experimentally  advantageous when $m_w^2\neq 0$.

Unfortunately, the index of refraction of normal matter for optical laser light is greater than $1$,
this means that $m_\gamma^2$ is negative (see Eq.~(\ref{index})) and the matching condition cannot be fulfilled.
Nevertheless this technique can still be useful, as proven by the ALPS collaboration~\cite{Ehret:2009sq,Ehret:2010mh}, to increase the
probability of oscillations to WISPs whose mass would be accidentally tuned to the experiment length $L$
such that the bracket in Eq.~(\ref{generalprob}) cancels.

The above picture is valid for any WISP having mass mixing with the photon.
In the ALP case only one photon polarisation (the one parallel to the magnetic field for parity-odd
couplings, and the one orthogonal for the parity even version) mixes and therefore only photons with this polarisation can convert into ALPs.
In the massive hidden photon case, each of the two photon polarisations mixes with the parallel hidden photon polarisation with identical strength (${\cal A}_0^{\gamma^\prime}= \chi \muu^2/(2 \omega)$, in this case).

This picture is illuminating but has its limitations.
One can imagine that photon-WISP conversions can happen back and forth many times within the length $L$ contributing to the final probability, as, e.g., in $\gamma\to \phi\to \gamma\to \phi\to ...\to \phi$.
Each conversion introduces a factor ${\cal A}_0^w$ and a length integral in the amplitude, so these processes are more and more suppressed if ${\cal A}_0^w L$ is small, which is the typical case we are interested in.
However they can be important in two circumstances: a) when  ${\cal A}_0^w L\gtrsim 1$ (what we will call the large mixing case) and b) when the computed probability in Eq.~(\ref{generalprob}) is zero (which happens typically when
$\Gamma_\gamma=\Gamma_w=0$ and $(m_w^2 -m_\gamma^2)L/(4\omega)$ is an integer multiple of $\pi$.

It would be very complex to sum these higher order transition amplitudes in the fashion outlined before
but fortunately there is an easy way out.
The key point is to recall that summing all the tree level diagrams (exactly what we want) is equivalent to solving the classical equations of motion~\cite{Raffelt:1987im}.
Since it will be of little practical use in this review, let us simply display the final result.
In the $\Gamma_\gamma=\Gamma_w=0$ limit one finds
\be
P(\gamma \to {\rm WISP} ) =
\frac{(2\delta m^2)^2}{(m_w^2-m_\gamma^2)^2+(2\delta m^2)^2}
\sin^2\( \frac{\sqrt{(m_w^2-m_\gamma^2)^2+(2\delta m^2)^2}L}{4\omega}\) ,
\ee
which has the same oscillatory behavior as Eq.~(\ref{Palp}) but with a correction to the oscillation frequency and amplitude involving $\delta m^2$.
The corrections are relevant only when  $\delta m^2 L/(2\omega)\gtrsim 1$, as expected.
Interestingly this conclusion holds even in the so-called resonant case ($m_w= m_\gamma$), where
$\delta m^2>|m_w- m_\gamma^2|$, because the $\sin$ can be Taylor expanded to recover the same coherent formula $(\delta m^2 L/\omega)^2$ that would be obtained with Eq.~(\ref{Palp}).
A detailed derivation for the ALP case can be obtained from \cite{Raffelt:1987im} for ultrarelativistic ALPs and from \cite{Adler:2008gk} for arbitrary values of $m_\phi$ (also in the context of LSW experiments).

Finally, let us emphasize that the WISP generation probability $P(\gamma \to {\rm WISP} )$ is identical
to the photon regeneration probability $P({\rm WISP}\to \gamma )$, for the same
external parameters such as length $L$ or magnetic field $B$.
Thus, we are now prepared to understand the predictions
of light shining through a wall via $\gamma\leftrightarrow {\rm WISP}$
oscillations and the corresponding experimental results.

\section{Laser light shining through a wall experiments}\label{sec:exper}

\begin{figure}[t!]
\begin{center}
\includegraphics[width=10cm]{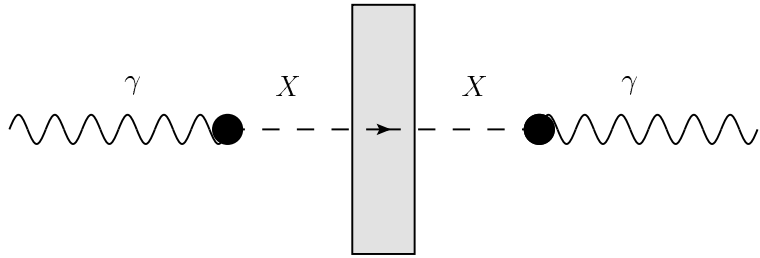}
 \end{center}
\caption{\base
Schematic of a light-shining-through a wall experiment.
}\label{fig:LSWillu}
\end{figure}

In a light shining through a wall experiment, the photon oscillates into
a WISP before the wall, the latter traverses the wall
and oscillates back into a photon behind the wall (see the illustration Fig.~\ref{fig:LSWillu}),
leading to a total LSW probability
\begin{equation}
\label{PLSW}
P_{\rm LSW} = P(\gamma \to {\rm WISP}; L_g, B_g, \ldots ) P({\rm WISP} \to \gamma ; L_r, B_r, \ldots ) ,
\end{equation}
where the conversion and back-conversion probabilities are given by Eq.~(\ref{generalprob}).
Therefore, one has to pay the price for the tiny conversion probability
twice.
%
%
\begin{figure}[tbp] 
  \centering
  \includegraphics[width=16cm]{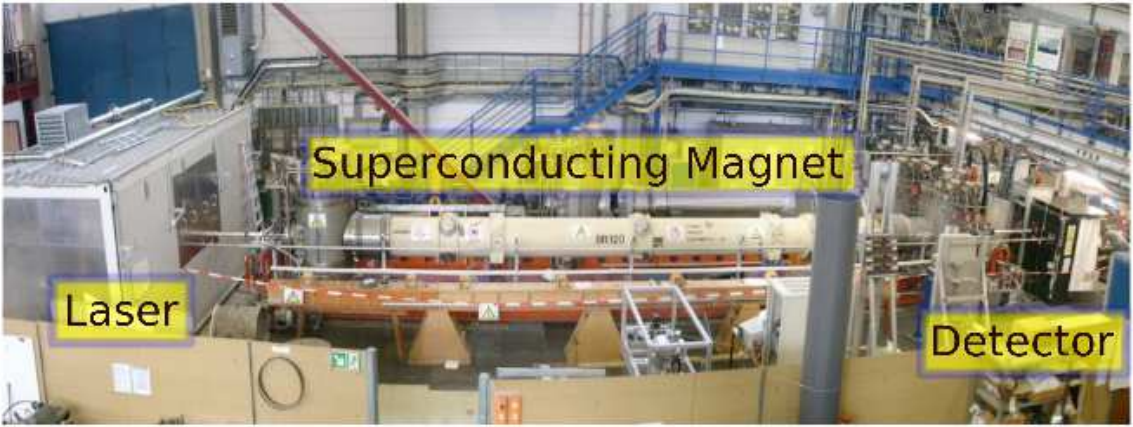}
  \caption{The ALPS experiment at DESY~\cite{Ehret:2009sq,Ehret:2010mh}.
  The primary laser system is kept in a light tight hut in the left hand side. From it, laser light is injected in the bore of
  a HERA superconducting dipole magnet. A blocking wall is placed in the center of
  the magnet's inner bore. Light eventually regenerated behind the wall would continue its trip through the magnet
  to end up in a black cabinet enclosing the photon detector.
  }
  \label{fig:ALPSexperiment}
\end{figure}
The number of photons detected after the wall in a general LSW experiment will then be given by
\be
N_{\rm LSW} = \underbrace{ (\beta_g {\cal P}_{\rm prim}/\omega )\Delta t}_{\rm \#\ of\ photons\ hitting\ wall}\ P_{\rm LSW}\ \beta_r\ \eta ,
\ee
where ${\cal P}_{\rm prim}$ is the primary laser power, $\Delta t$ is the time it is switched on providing photons,
$\omega$ is the photon energy,
and $\eta$ stands for the efficiency of collecting and detecting the photons regenerated after the wall. The
two additional factors, $\beta_{g,r}$ account for possible enhancements of the signal if one exploits
different experimental tricks to recycle the photons, i.e. if every photon is driven against the wall or the
detector a number of times.
For instance, if one uses an optical resonant cavity to trap the photons in the WISP generation part of the experiment the
factor, $\beta_g$ is given by the power buildup of the cavity.
In principle, a similar trick is possible in the photon regeneration part of the experiment, so we have included $\beta_r$ for
completeness (see Section~\ref{sec:next}).

\begin{table*}
\begin{center}
\begin{tabular}{|l|c|c|c|c|}
\hline\small
Experiment &\small $\omega$ &{\small $P_g$}& $\beta_g$ &\small Magnets \\
\hline
{\small ALPS} (DESY)~\cite{Ehret:2009sq,Ehret:2010mh} & 2.33~eV   & 4~W&300&\begin{minipage}[c]{4.cm}\centering $\MF_g=\MF_r=5$~T\\
$L_g=L_r=4.21$~m\end{minipage}  \\[0.1cm]
\hline
{\small BFRT} (Brookhaven)~\cite{Ruoso:1992nx,Cameron:1993mr} & 2.47~eV & 3~W&100&
\begin{minipage}[c]{4.cm}\centering $\MF_g=\MF_r=3.7$~T\\ $L_g=L_r=4.4$~m\end{minipage}  \\[0.1cm]
\hline
{\small BMV} (LULI)~\cite{Fouche:2008jk,Robilliard:2007bq} & 1.17~eV&$8\times10^{21}$ $\frac{\gamma}{\rm pulse}$ (14~pulses) &1&
\begin{minipage}[c]{4.cm}\centering $\MF_g=\MF_r=12.3$~T\\ $L_g=L_r=0.4$~m\end{minipage}  \\[0.1cm]
\hline
{\small GammeV} (Fermilab)~\cite{Chou:2007zzc} & 2.33~eV    &$4\times10^{17}$ $\frac{\gamma}{\rm pulse} $ (3600~pulses) & 1  &
\begin{minipage}[c]{4.cm}\centering $\MF_g=\MF_r=5$~T\\ $L_g=L_r=3$~m\end{minipage}  \\[0.1cm]
\hline
{\small LIPSS} (JLab)~\cite{Afanasev:2008jt,Afanasev:2008fv} &1.03~eV &$180$~W    &1&
\begin{minipage}[c]{4.cm}\centering $\MF_g=\MF_r=1.7$~T\\ $L_g=L_r=1$~m\end{minipage}  \\[0.1cm]
\hline
{\small OSQAR} (CERN)~\cite{Pugnat:2007nu,Ballou} &2.5~eV &15~W &1&
\begin{minipage}[c]{4.cm}\centering$\MF_g=\MF_r=9$~T\\$L_g=L_r=7$~m\end{minipage} \\[0.1cm]
\hline
{\small BMV} (ESRF)~\cite{Battesti:2010dm} &50/90~keV &10/0.5~mW &1&
\begin{minipage}[c]{4.cm}\centering$\MF_g=\MF_r=3$~T\\$L_g=1.5, L_r\sim 1$~m\end{minipage} \\[0.1cm]
\hline
\end{tabular}
\vspace{2ex}
\caption{Some experimental parameters of the past and current generation of LSW experiments.}
\label{tab:LSWexp}
\end{center}
\end{table*}

Clearly, one has to exploit the highest photon fluxes
and the most sensitive photon detectors in order to be sensitive to
very small LSW probabilities, maximizing the WISP discovery potential.
For this reason, all of the present LSW experiments employ lasers in the optical regime (see Table~\ref{tab:LSWexp}).
Those deliver currently the photon beams with the highest flux and best coherence properties.
In fact, commercial lasers in the visible spectrum can easily reach output powers of several tens of Watts, corresponding to $\gtrsim 10^{19}$ photons per second.
The highest photon flux, currently, has been exploited by the ALPS (Any Light Particle Search) experiment~\cite{Ehret:2009sq,Ehret:2010mh} at DESY (cf. Fig.~\ref{fig:ALPSexperiment})
which is the first one to exploit a resonant optical cavity on the emitter side of the apparatus to increase the emission probability by a huge factor. All the other experiments so far exploited multi-pass delay lines  or operated in a single pass mode.
Sensitive detectors with quantum efficiencies close to $\sim 100\%$ in the optical spectral
range are also available.
Finally, a very important point is the suppression of backgrounds in the detection process. Ambient light, cosmic rays, environmental radioactivity or even the read-out electronics of the detector can produce
fake LSW events that have to be studied and discriminated from real LSW photons.
There are different approaches to this issue as we shall see later.

The third main ingredient for an all-round LSW experiment are strong magnets which are
otherwise exploited typically in particle accelerators. Correspondingly, almost all the LSW
experiments are located in accelerator laboratories, such as CERN, DESY, Fermilab and Jefferson Lab
(see Table~\ref{tab:LSWexp}).

But before we discuss the current results of LSW experiments, let us recapitulate the historical
development of this field of research.

\subsection{Historical development}

The first proposal for a light shining through a wall experiment
appeared in 1982~\cite{Okun:1982xi} as a proposal to search for hidden photons.
In the context of axion-like particles, such experiments were proposed three years later~\cite{Anselm:1986gz,VanBibber:1987rq}.
In the early 1990s these proposals were realized by the so called BFRT collaboration, named this way after the four parties involved: Brookhaven National Laboratory, Fermilab, Rochester and Trieste universities~\cite{Ruoso:1992nx,Cameron:1993mr}.
They performed pioneering experiments not only of the light shining through walls type but also searched for changes in laser polarization.
The core of the experiment was the use of recycled dipole magnets from the colliding beam accelerator (CBA) at
Brookhaven National Laboratories.
No light shining through walls was observed and, accordingly, the limits on the conversion probability were used to constrain the parameters of axion-like particles and hidden photons~\cite{Ruoso:1992nx,Cameron:1993mr}.

Once the experiment was finished, the Italian part of the collaboration continued the development of the techniques to improve the sensitivity for laser polarisation experiments building their own experiment at the INFN Legnaro in Italy.
In 2005, this so called PVLAS (acronym for `Polarization of the Vacuum with a LASer') experiment reported the
observation of an anomalously large rotation of the polarisation plane of photons after the passage through a magnetic field~\cite{Zavattini:2005tm}.
This result provided the impetus for a number of new laser light shining through a wall (LSW)
experiments (cf. Table~\ref{tab:LSWexp}) to search for
photon $\to$ WISP $\to$ photon conversions rather than solely for disappearance.
This new generation of LSW experiments
could improve the constraints from the pioneering experiment BFRT~\cite{Cameron:1993mr} by about an order of magnitude in the WISP--photon coupling (cf. Fig.~\ref{fig:LSWresult})~\cite{Robilliard:2007bq,Chou:2007zzc,Pugnat:2007nu,Ballou,Afanasev:2008jt,Fouche:2008jk,Afanasev:2008fv,Ehret:2009sq,Ehret:2010mh}.
Moreover, the momentum gained by these experiments towards the establishment of a new
low-energy, high intensity frontier of particle physics turned out to be conserved even though the original motivation disappeared: the PVLAS collaboration could not confirm their first observation after an upgrade of their apparatus~\cite{Zavattini:2007ee}. This is in-line with the finding of the above
mentioned LSW experiments.

Let us describe next the current status of the LSW experimental programme and its outcome in some
detail\footnote[13]{We will concentrate on laser based LSW experiments which are currently quite mature.
Analogous microwave cavity based ``L"SW experiments~\cite{Hoogeveen:1992uk,Jaeckel:2007ch,Caspers:2009cj},
are still in their infancy~\cite{Povey:2010hs,Wagner:2010mi}.}
.

\subsection{Current status}

The experiments in Table~\ref{tab:LSWexp} were running in quite a number of different setups.
Data sets were taken with magnet on or off, laser polarization parallel or perpendicular to the magnetic field, and different
(rest) gas pressures in the production/regeneration tubes, corresponding to different refractive indices, $n$,
in order to get as much information as possible.
The experiments were designed to survey the region of parameter space favored from the PVLAS signal, which was determined to be~\cite{Ahlers:2006iz},
\be
m_\phi\sim {\rm meV}\; \; \; ; \; \; \; g \sim 2-3\times 10^{-6}\ {\rm GeV}^{-1},
\ee
in the case of an ALP and $m_{\rm MCP}\lesssim 0.1$ eV, $Q\sim 10^{-6}$ in the case of an MCP.

The BMV experiment was designed in order to profit from very intense pulsed magnetic fields and lasers.
The experiment was setup in the Laboratoire pour l'Utilisation des Lasers Intenses (LULI), in the Nano 2000 chain,
where 1050 nm laser light pulses up to 1.5 kJ (in 4.8 ns) can be delivered at a repetition rate of 1 pulse per two hours.
Since the laser was pulsed, the regenerated photons had to arrive to the detector in a very tiny time window, where
dark count rates are typically negligible.
Emphasis was then put into obtaining a very good single photon detection efficiency
($\sim$ 50\%). The dark count rate during a pulse was extremely small, $\sim 5\times 10^{-4}$.
Finally, two pulsed magnets (producing more than 12.3 T during 150 $\mu$s) delivered from LNCMP
were used for the generation of ALPs and regeneration of photons.
The magnetic field produced was not constant along the photon's trayectory and the conversion/reconversion probability had to be computed directly from the phase integral in Eq.~(\ref{eq:ALPoscillationintegral}) by using $B_{\rm ext}=B_{\rm ext}(z)$. The effective length of the magnetic field was $\sim 0.37$ m.
With only 14 laser pulses they were able to rule out the PVLAS favored region, since no regenerated photon
was detected~\cite{Fouche:2008jk,Robilliard:2007bq}.

The GammeV experiment used very much the same concept, except that in its case, they exploited a single dipole from the
Tevatron accelerator (6 m length and 5 T field strength) instead of pulsed magnets. As a light source they used a
Nd:YAG laser delivering 150 mJ of 532 nm light in 5 ns pulses.
The much lower laser intensity required a bigger number of pulses to exclude the PVLAS ALP, $\sim 3600$,
but the higher repetition rate, 20 Hz, made this possible in 20 h of data taking.
The long and strong Tevatron magnets had an unfortunate length, somehow tuned to the oscillation length
of the hypothetical PVLAS ALP (using half of the magnet for ALP generation and the other for
photon regeneration was producing a dip in sensitivity at $m_\phi\sim 1.4$ meV).
In order to fill this dip, the GammeV experiment took data also in an asymmetric configuration $L_g=1$ m,
$L_r=5$ m. Taking 20 hours of data in each of these configurations, and with two different laser polarizations
(parallel and perpendicular to the magnetic field) they were able to exclude the PVLAS ALP regardless of its
scalar/pseudoscalar nature~\cite{Chou:2007zzc}.

The ALPS experiment at DESY (cf. Fig.~\ref{fig:ALPSexperiment})  fancied a  8.8 m long, 5 T magnetic field provided by one
dipole magnet designed for the HERA accelerator~\cite{Ehret:2009sq,Ehret:2010mh}. The photon source was a MOPA laser system producing up to 35 W of 1064 nm laser light, which was frequency doubled to 532 nm
($\omega=$2.33 eV) to optimize the detection efficiency.
Instead of triggering the photon detector to record data only in coincidence with laser pulses they collected
continuously the possible LSW signal by focusing the hypothetical beam of regenerated photons in a few pixels of
a CCD camera. This technique is certainly challenging, but can be scaled very easily to more sensitive experiments.
Since the measurement time is much longer (ALPS analyzed dozens of hours of data, while for instance BMV analyzed 14 pulses of 5 ns!)
the backgrounds will be enormous, thermal emission from the black box encompassing the camera, cosmic rays, environmental radioactivity or even the thermal activity of the detector produce eventually fake events, i.e. will be collected and understood by the detector as if it were LSW. All these effects have to be minimised.
Focusing the signal in the smallest collecting area is essential as all the background events are in principle randomly distributed and therefore diminish linearly with the exposed area. Moreover, the rest of the CCD pixels can be used to study and characterize the backgrounds.
Thermally induced events from the environment or the camera itself can be reduced by cooling down the system.
For instance ALPS operated its CCD at -70 $^{\circ}$C and had a dark count rate of $10^{-3}$ counts/pixel/s.
The quantum efficiency at 532 nm was $96\%$.
Cosmic rays or radioactivity are more difficult to avoid and they impose a serious restriction on the operation of a LSW experiment.
They typically deposit a lot of energy on the detector and in a extended area leaving traces which are easily recognizable, so they can be rejected by inspection with the naked eye or computer algorithms.
Unfortunately, if one measures long enough eventually one of these traces will appear in the few pixels where the
LSW signal is expected spoiling the measurement. Therefore in practice one proceeds not with a single measurement run but splits it in smaller runs, typically of $\sim$1 hour each, and discards those runs in which large energy depositions appear
in or near the signal region.

The most remarkable aspect of the experiment was the setup of an optical cavity
in the ALP generation part of the experiment. Keeping the cavity on resonance, the laser power
was boosted inside the cavity up to 1200 Watt while fed by 4.6 W of 532 nm laser light.
As we will comment later on, the use of optical cavities both in the generation and regeneration
parts of a LSW experiment seems to be the unique path towards a much more sensitive
new generation of LSW experiments so in this sense, ALPS was a very important pioneer experiment.
The ALPS results imply nowadays the most stringent purely laboratory constraints on different WISPs such as ALPs, MCPs and hidden photos, excluding the PVLAS interpretation in terms of ALPs or MCPs.

The LIPSS experiment was performed at Jefferson Lab~\cite{Afanasev:2008jt} and benefited from extremely intense pulses of laser
light provided by their free-electron laser (180 Watt of average power in 150 fm pulses at a 75 MHz repetition rate).
LIPSS used $1$ m long, 1.77 T dipole magnets. They focused their search on scalar ALPs (the laser polarisation was perpendicular to the direction of the magnetic field)
and on hidden photons~\cite{Afanasev:2008jt,Afanasev:2008fv}.

The OSQAR experiment at CERN, although mainly devoted to the experimental test of QED vacuum magnetic birefringence~\cite{Heisenberg:1935qt}, performed a dedicated experiment to test the ALP interpretation of the PVLAS signal.
The main feature of the experiment is the use of one of the longest and strongest dipole magnets available,
those engineered for the Large Hadron Collider, with a length of 14.3 m and field strength of 9 T.
As the photon source they used a ionized Argon laser delivering up to 18 Watt of power in multiwavelength mode
(514 and 488 nm, corresponding to 2.41 and 2.54 eV are the most important frequencies) and as a detector a
liquid nitrogen cooled CCD with quantum efficiency $\sim 50\%$ at the relevant wavelengths and a dark count rate smaller than
$0.1$ counts/pixel/min~\cite{Pugnat:2007nu,Ballou}.

However, a note is in order concerning the ALP interpretation of the
measurements of OSQAR. To this end,
we should recall the $\gamma\to$ ALP conversion probability formula, Eq.~(\ref{Palp}).
For $m_\phi\sim$ meV, $\omega \sim 2.5$ eV and $L\sim 7$ m the phase of the sinus is $\sim \pi$ and
therefore probability is suppressed.
In order to get the most of the LHC dipole, the OSQAR collaboration planned to fill the oscillation region with
a suitable gas restore the coherence of the $\gamma\to$ALP conversions along the whole experiment path,
as described in Sec.~\ref{sec:oscillations}.
However, supported by a longstanding controversy about the momentum of photons in a medium\footnote[14]{
Meanwhile, the controversy has been very much clarified~\cite{Barnett:2010zz}.}, they interpreted
the photon mass $m_\gamma^2$ as $\omega^2- p^2$ with $p$ the physical momentum of a photon, and not
its wavenumber $k$, the relevant quantity to compute phase differences. Both magnitudes are related to
the index of refraction as $k=\omega n$, $p=\omega/n$ so the confusion amounts to take
$m_\gamma^2={\cal R}\{2\omega^2(n-1)\}$, i.e. with opposite sign with respect to Eq.~(\ref{index})
in Sec.~\ref{sec:oscillations}. Therefore, when they filled the oscillation region with N$_2$ gas to
make $m_\gamma^2\sim$ (meV)$^2$ with the intention to cancel the phase difference, what they
arranged for in reality was $m_\gamma^2\sim$ -(meV)$^2$, so that the phase of Eq.~(\ref{Palp}) was actually $\sim 2\pi$ and again suppressed.

\begin{figure}[t!]
\subfigure[]{\includegraphics[width=.4\textwidth,angle=270]{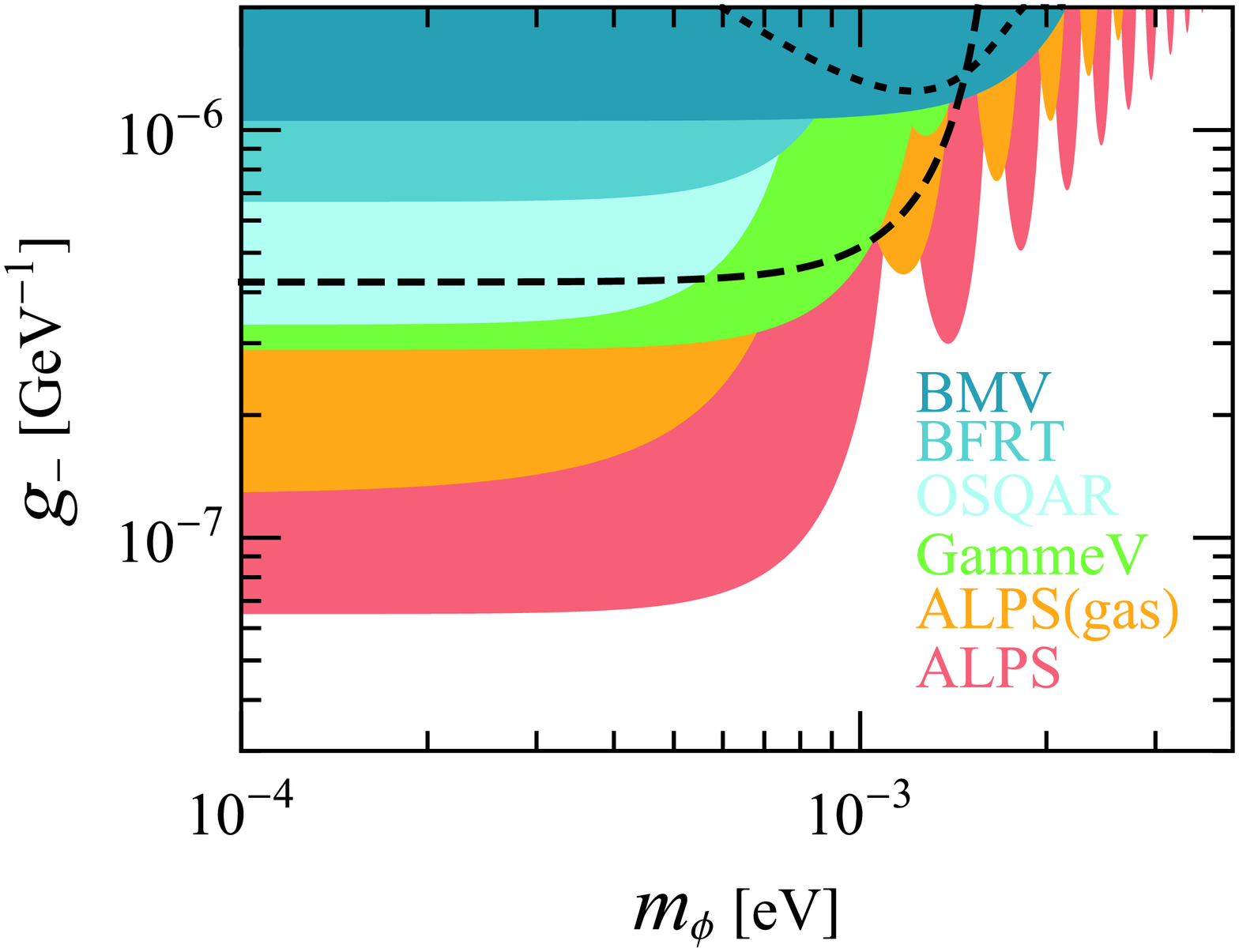}}
\hfill
\subfigure[]{\includegraphics[width=.4\textwidth,angle=270]{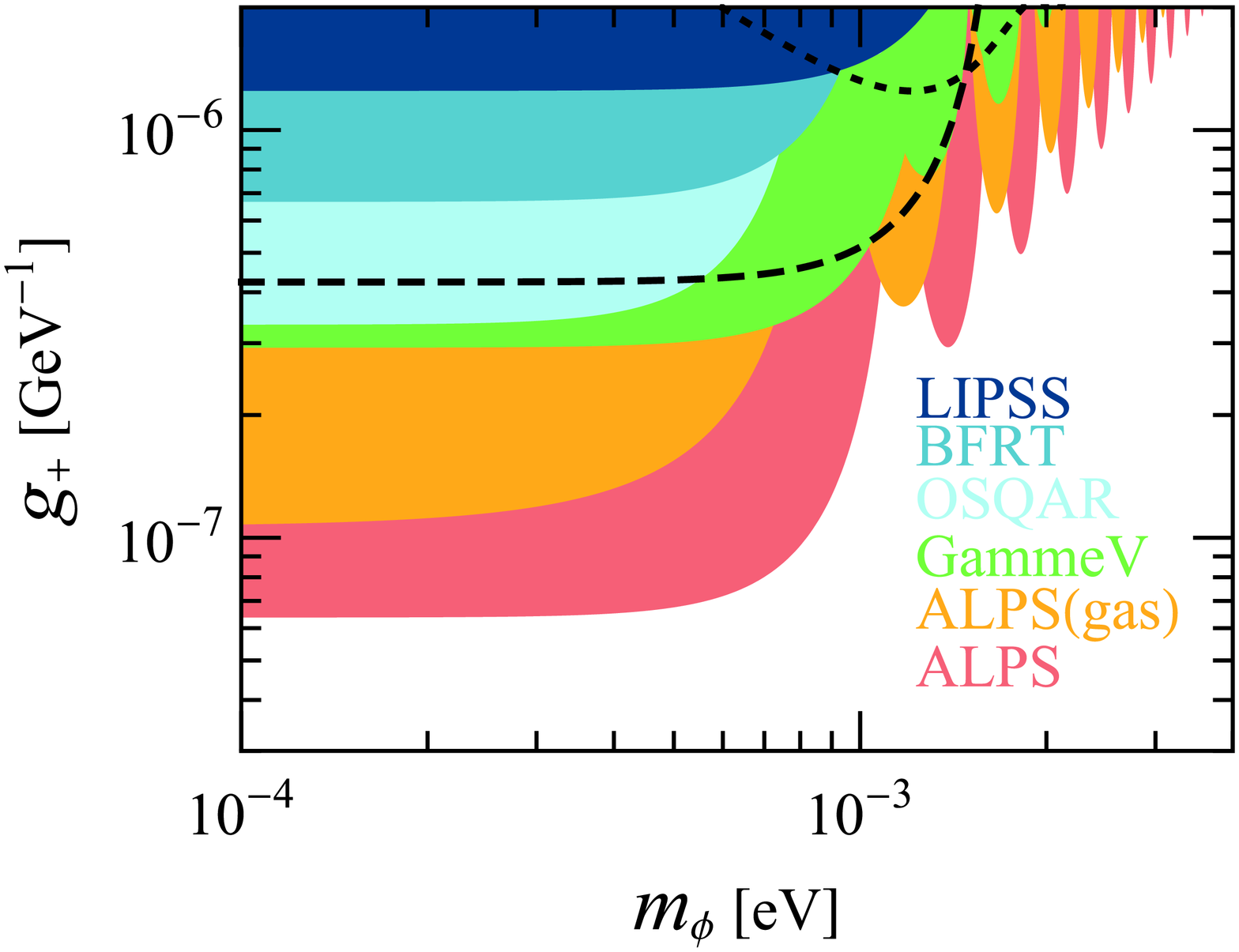}}
\subfigure[]{\includegraphics[width=.4\textwidth,angle=270]{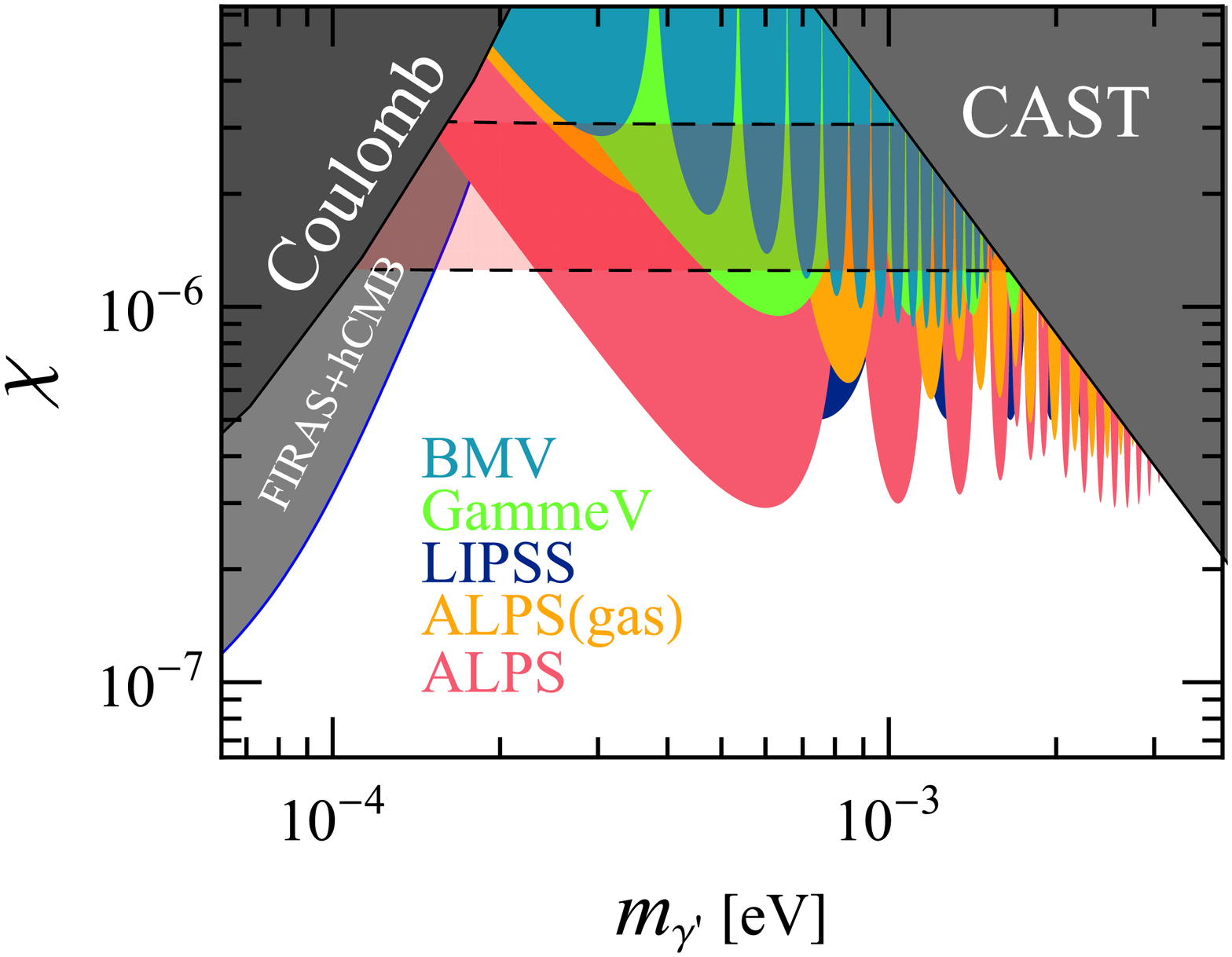}}
\hfill
\subfigure[]{\includegraphics[width=.4\textwidth,angle=270]{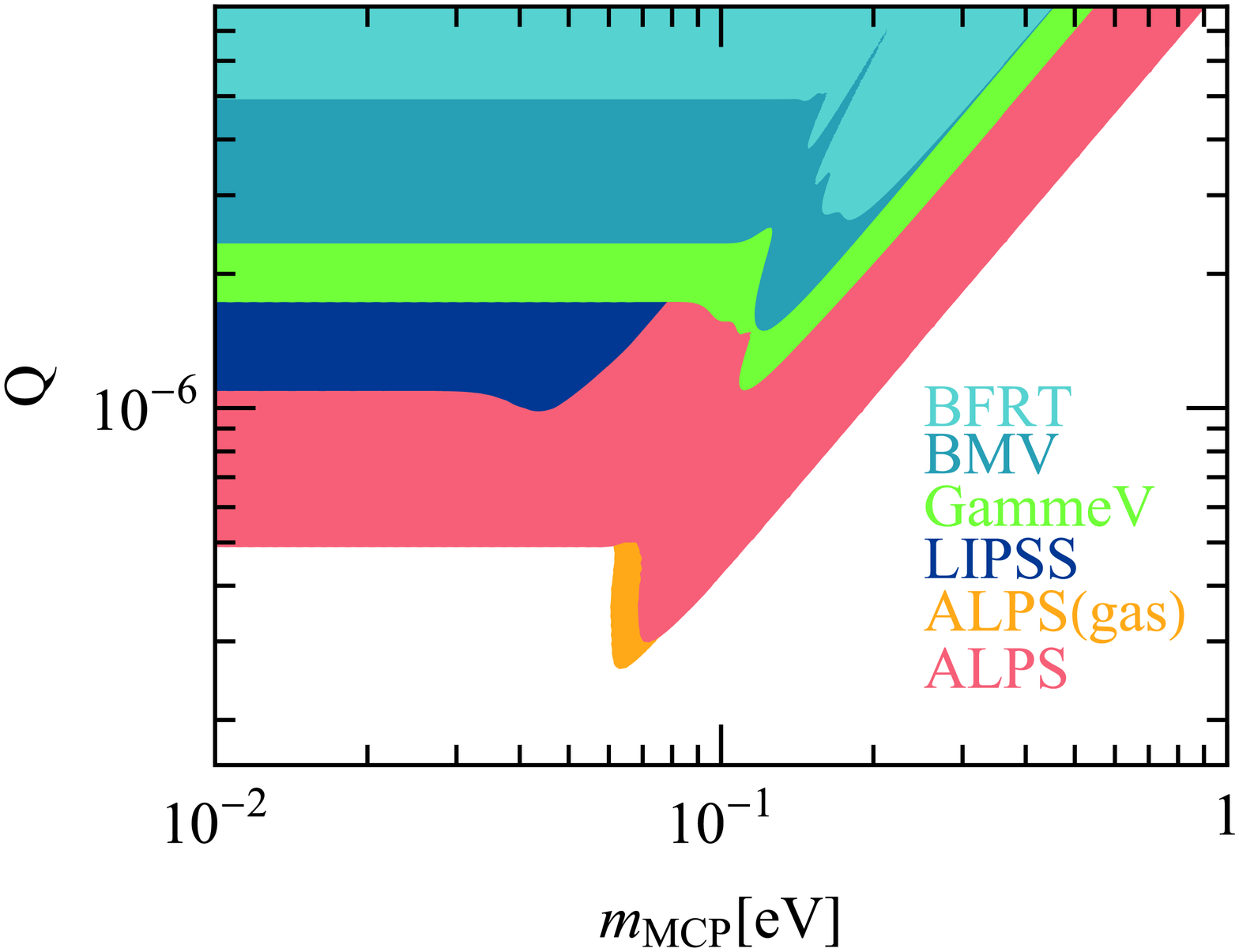}}
\caption{
Exclusion limits ($95\%$ C.L.) for WISPs from  from the LSW experiments
ALPS~\cite{Ehret:2010mh}, BMV~\cite{Fouche:2008jk,Robilliard:2007bq}, BFRT~\cite{Cameron:1993mr},
GammeV~\cite{Chou:2007zzc}, LIPSS~\cite{Afanasev:2008jt}, and OSQAR~\cite{Pugnat:2007nu,Ballou} (vacuum measurement). Top panels:
pseudoscalar (a) and scalar (b) axion-like particles.
Bottom panels: massive hidden photons (c) and massless hidden
photons with an additional minicharged particle (d).
Also shown, in panel (c), are limits from searches of modifications of Coulomb's law~\cite{Bartlett:1988yy},
distortions of the CMB spectrum~\cite{Jaeckel:2008fi} and the solar axion search by CAST~\cite{Redondo:2008aa}.
Hidden photons in the horizontal  band in panel (c) could account for the apparent excess in the relic
neutrino density recently reported by WMAP-7~\cite{Komatsu:2010fb}.
Compilation from Ref.~\cite{Ehret:2010mh}.}
\label{fig:LSWresult}
\end{figure}

Despite the effort of the above mentioned collaborations unfortunately no light shining
through a wall was detected and the PVLAS ALP interpretation was independently excluded.
Correspondingly, the experiments published experimental upper limits on the probability for light shining through a wall.
A comparison with the corresponding prediction in the context of one of the WISP models
yields then the corresponding upper limit on the WISP coupling constant vs. its mass.
A graphical summary of the exclusion bounds achieved is presented in Fig.~\ref{fig:LSWresult}.

Let us now spend a few words on the current results for the different WISP candidates.

Pseudoscalar ALPs couple to two photons through Eq.~(\ref{twogammacoupl}).
Correspondingly, $\gamma\to$ALP conversions may occur only if the external magnetic field is
switched on and the photon polarization is parallel to it (henceforth $\gamma_{||}$).
The conversion probability is predicted as (see last section)
\begin{eqnarray}
\label{axionprob}
P(\gamma_{||} \leftrightarrow \phi^{(-)}  )= 4\frac{(g_{_-} \omega B)^2}{\(m_\phi^2+2\omega^2(n-1)\)^2}
\sin^2\left(\frac{m_\phi^2+2\omega^2(n-1)}{4\omega}L_B\right),
\end{eqnarray}
with $B$ the magnitude of the magnetic field orthogonal to the photon's direction of motion and $L_B$ its length.
For a scalar ALP $\phi^{(+)}$, $g_- $ has to be replaced by $g_+$
in Eq.~(\ref{axionprob}) and $\gamma_{||}$ by $\gamma_{\perp}$, i.e. this time, only photons
polarized perpendicularly to the external magnetic field can convert into ALPs and viceversa.
With the help of Eq.~(\ref{axionprob}) it is easy to translate the limits on the probabilities
of light shining through a wall into limits of the couplings
$g_\pm$ vs. ALP mass $m_\phi$, cf. Fig.~\ref{fig:LSWresult} (a) and (b).

The most stringent constraints on $g_\pm$ are generally obtained for massless ALPs for vacuum conditions in the
beam pipes ($n\equiv 1$) since generally $n>1$ for (near) visible light and the photon effective mass $2\omega^2(n-1)$ suppresses the conversions.
For larger masses, the $\gamma\to$ALP conversions increasingly lose coherence and there are even regions in coupling vs. mass which are unconstrained by vacuum measurements.
These regions correspond especially to masses for which $m_\phi^2 l/(4 \omega)=\pi\times {\rm integer}$.
These gaps in sensitivity can be filled by introducing an adequate amount of gas in the conversion and reconversion regions such
that $2\omega^2(n-1)L/(4 \omega)=\pi/2$, making the $\sin$ in Eq.~(\ref{axionprob}) equal to one.
For instance, in the ALPS setup, $L=4.3$ m and $\omega=2.33$~eV, so the above condition was
achieved by introducing Ar gas at a pressure of $0.18$~mbar so that $(n-1)\simeq 6.2 \times 10^{-8}$.

The oscillations into  hidden photons occur also in the absence of a magnetic field (cf. Fig.~\ref{fig:LSWcand} (b)).
The conversion probability is predicted as~\cite{Ahlers:2007qf}
\begin{equation}
\label{hpprob}
P(\gamma \leftrightarrow \gamma^\prime  ) \simeq
4\chi^2 \frac{m_{\gamma^\prime}^4}{\(m_{\gamma^\prime}^2+2\omega^2(n-1)\)^2}\sin^2\left(\frac{m_{\gamma^\prime}^2+2\omega^2(n-1)}{4\omega}L\right) ,
\end{equation}
where $L$ is now the propagation length.
Clearly, the conversion probability vanishes for $m_{\gamma^\prime}\to 0$. This is
also apparent in Fig.~\ref{fig:LSWresult} (c) which displays the upper bound on the kinetic mixing $\chi$ vs. $m_{\gamma^\prime}$.

Finally, even in the $m_{\gamma^\prime}=0$ case, $\gamma\to \gamma'$ oscillations are possible in a magnetic field if there are light particles charged under the hidden U(1), i.e. mini-charged particles (cf. Fig.~\ref{fig:LSWcand} (c)).
Since the corresponding probability is too involved we refer the reader to~\cite{Ahlers:2007rd,Burrage:2009yz}.
The limits on the charge of mini-charged particles vs. their mass are displayed in Fig.~\ref{fig:LSWresult} (d),
where we have assumed a Dirac MCP and $e_h=e$ for
simplicity\footnote[15]{For smaller values of $e_h$, like the ones arising in
hyperweak scenarios~\cite{Goodsell:2009xc}, all laboratory bounds get worse, see Ref.~\cite{Ahlers:2007qf}}.

In summary, LSW experiments have improved their sensitivity for WISPs considerably in recent years.
In fact, for those WISPs which couple to photons they deliver meanwhile the tightest, purely laboratory-based
constraints. To put these bounds in a global perspective, however, we will
review in the next section also other -- often stronger, but more model-dependent -- constraints,
which arise in the context of astrophysics and cosmology.

\section{WISPs in astrophysics and cosmology: light shining through other stuff\label{sec:astrocosm} }

The phenomenon of light shining through walls via an intermediate WISP state can have very important
consequences in environments other than our laboratories.
Actually, the strongest bounds on the existence of WISPs presently often come from the non observation of these consequences
in stellar evolution, big bang nucleosynthesis, and the cosmic microwave background.
However, there are also some intriguing astronomical observations
which are hard to  explain by known physics and might be interpreted as indirect hints pointing towards
the existence of WISPs. In this section we briefly review these arguments.

\begin{figure}[t!]
\centerline{\includegraphics[width=.6\textwidth]{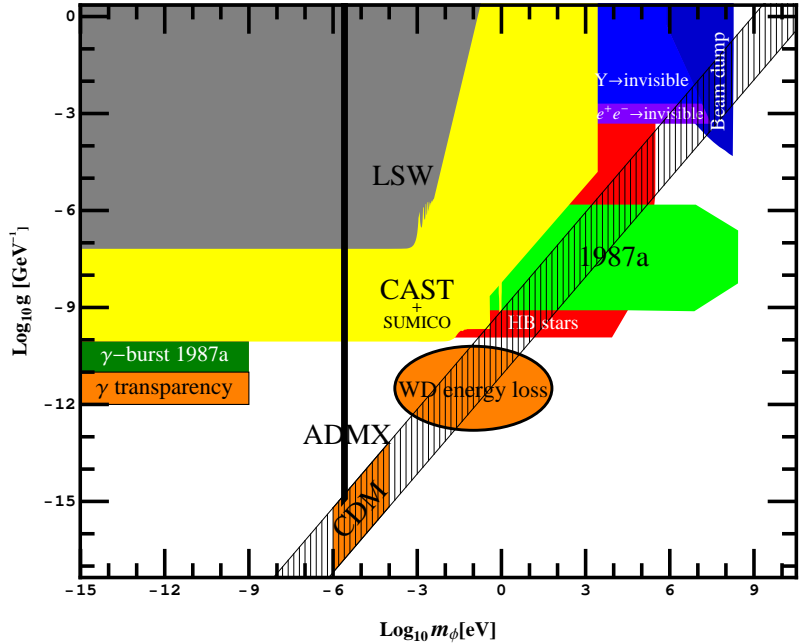}}
\caption{Summary of astrophysical, cosmological and laboratory constraints
on axions and axion-like-particles (two
photon coupling $g$ vs. mass $m_{a}$ of the
ALP). The hatched band
shows the theoretical prediction, Eqs.~(\ref{axionmass}) and (\ref{axioncoupling}),
for the QCD axion.
Two areas with special interest (not constraints) are shown in orange: The range where the axion can
be the cold dark matter (the orange region labeled ``CDM" in the
plot, which can be extended towards smaller masses by anthropic reasoning) and the range where
axions could explain the recently reported anomalous cooling of white dwarf stars (labelled WD energy loss).
For comparision, we also show laboratory limits from photon regeneration
experiments (ADMX and LSW).
Note that the limit from ADMX is valid only under the assumption that the
local density of ALPs at earth is given by the dark matter density.
Compilation from Ref.~\cite{Jaeckel:2010ni}.}
\label{fig:alps_astro}
\end{figure}

\subsection{Bounds from stellar evolution and helioscope searches}\label{stellar}

Just as laser light can make it through a wall by converting to WISPs in our labs,
photons present in the interior of stars can make it through the star's envelopes and escape as WISPs.
The ``invisible" energy loss implied by the emission of WISPs has to be provided by the
nuclear reactions powering the star and therefore constitutes a mismatch between the rate
at which the nuclear fuel is consumed and the standard stellar energy loss mechanisms: photons from the surface
and neutrinos from the stellar cores. This mismatch can be constrained by comparing
observations with numerical simulations of stellar evolution, and leads to very stringent constrains
of the interactions of WISPs~\cite{Raffelt:1996wa}.

\begin{figure}[t!]
\centerline{
\includegraphics[angle=0,width=.7\textwidth]{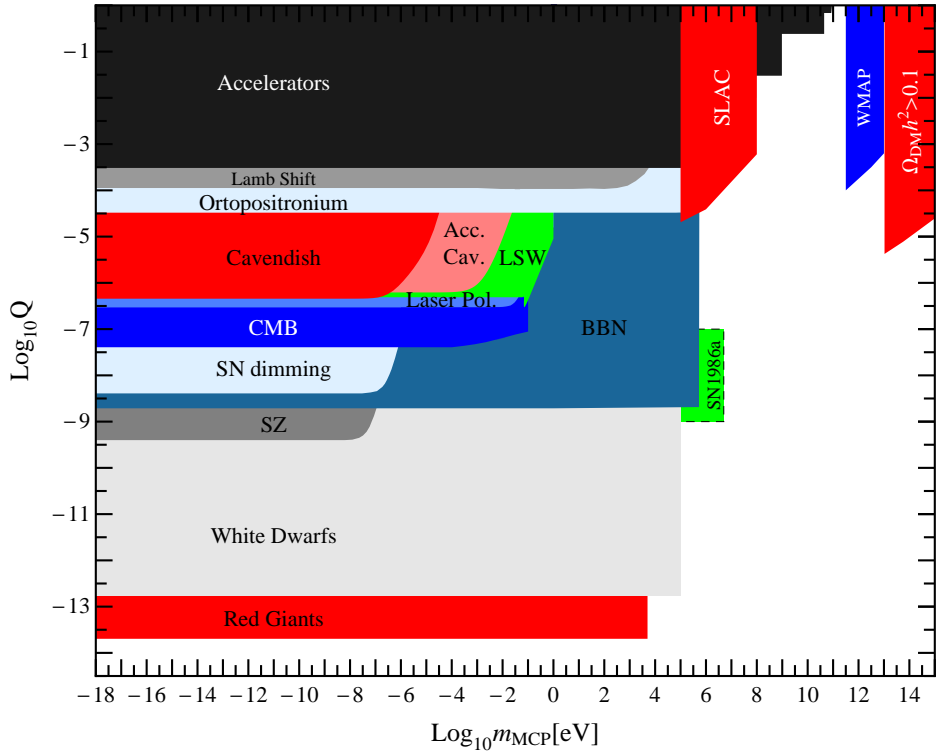}
}
\caption{
Summary of astrophysical, cosmological and laboratory constraints on minicharged particles (fractional charge
$Q$ vs. mass $m_{\rm MCP}$).
At relatively large
masses and couplings we also have the bounds from accelerator and fixed target experiments (SLAC).
Compilation from Ref.~\cite{Jaeckel:2010ni}.}
\label{fig:mcp_astro}
\end{figure}

The strongest limits for general ALPs with a two photon coupling and MCPs come from observations
of Horizontal Branch (HB) stars in globular clusters~\cite{Raffelt:1985nk,Raffelt:1987yu}
(cf. Figs.~\ref{fig:alps_astro} and \ref{fig:mcp_astro}). For very small masses an even tighter limit on a two-photon coupling of ALPs
can be obtained from the absence of a $\gamma$-ray burst in coincidence with a neutrino burst during the
supernova explosion SN 1987a~\cite{Brockway:1996yr}.
The principle behind the latter bound is that ALPs would be produced in the supernova core from the Primakoff effect and reconverted into $\gamma$-rays inside the galactic magnetic field.

Interestingly, a possible non-standard energy loss has been recently identified in the white dwarf (WD) luminosity function~\cite{Isern:2008nt}.
As pointed out by the authors this is compatible with the existence of ALPs with an ALP-electron coupling,
$g_{ee\phi}\simeq 10^{-13}$,
suggesting a decay constant $f_\phi$ and corresponding coupling to photons $g$ of order
\begin{equation}
f_\phi\sim g_{ee\phi} m_e \sim  {\rm few}\times 10^{9}\ {\rm GeV}\hspace{6ex} \Rightarrow
\hspace{6ex}
g\sim \alpha/f_\phi  \sim 10^{-12}\ {\rm GeV}^{-1},
\label{fa_benchmark_axion_wd}
\end{equation}
respectively.
The latter is quite close to the stellar evolution bounds (cf. Fig.~\ref{fig:alps_astro}).
Obviously, it is quite possible that
a more conventional explanation for this non-standard energy loss may be found.
Fortunately, the corresponding region in parameter space may be eventually checked in a future LSW experiment.
Finally, it is also worth nothing that very similar ALPs have been invoked to solve some problematic aspects of the X-ray activity of the Sun,
the longstanding corona problem and the triggering of solar flares~\cite{Zioutas:2009bw}.

\begin{figure}[t!]
\centerline{
\includegraphics[width=.85\textwidth]{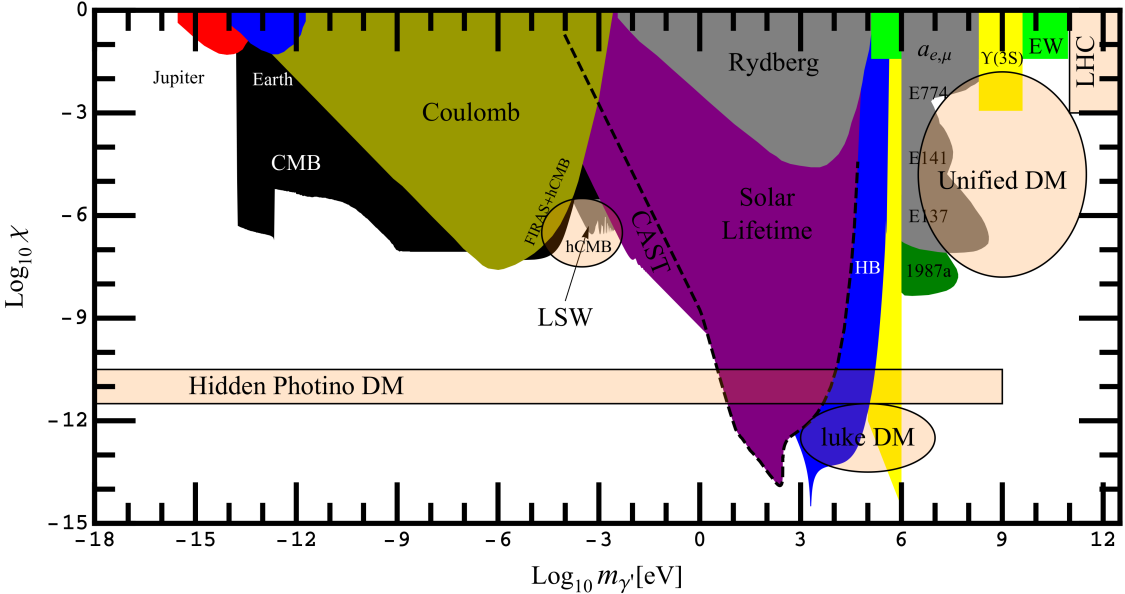} 			}
\caption{
Summary of astrophysical, cosmological and laboratory constraints for hidden photons (kinetic mixing
$\chi$ vs. mass $m_{\gamma^\prime}$).
At higher mass we have electroweak precision measurements (EW), bounds from upsilon decays ($\Upsilon_{3S}$)
and fixed target experiments (EXXX)). Areas that are especially interesting are marked in light orange.
Compilation from Ref.~\cite{Jaeckel:2010ni}.}
\label{fig:hp_astro}
\end{figure}

The Sun is less sensitive (even though its properties are better known)
than these other stars to axion or MCP emission, since its inner density and temperature are a bit smaller, and so it would be the WISP emission.
Solar bounds have been obtained from studies of its
lifetime, helioseismology and the neutrino flux~\cite{Schlattl:1998fz,Gondolo:2008dd}, but although the data is more precise the resulting constraints are weaker.
However, as is apparent in Fig.~\ref{fig:hp_astro}, this is different for hidden photons:
the region in parameter space excluded by the solar lifetime~\cite{Redondo:2008aa} complements
in this case the one excluded by the lifetime of HB stars~\cite{Redondo:2008ec}.

Instead of looking for the indirect effects due to WISP production in the sun, e.g. its reduced lifetime,
{\em helioscopes} try to detect the WISPs directly on earth~\cite{Sikivie:1983ip,vanBibber:1988ge,Gninenko:2008pz}.
Basically, they employ the same idea as an LSW experiment, the difference being that the laboratory-bound
WISP generation side is replaced by solar WISP generation in the sun's interior. The wall is simply everything
in between the solar core and the regeneration side (the rest of the sun, the atmosphere, the walls of the
experimental hall etc.). The latter has to be
pointed, of course, towards the sun.
The enormous total number of interactions inside the sun would lead to a huge WISP flux even if the coupling is tiny. This
makes helioscopes extremely powerful tools to search for WISPs, however, with some model dependence: if somehow
the production of WISPs inside the sun is suppressed, helioscopes loose their sensitivity~\cite{Jaeckel:2006xm}.

In fact, currently two axion helioscopes are running, CAST~\cite{Andriamonje:2007ew,Arik:2008mq} and SUMICO~\cite{Inoue:2008zp}.
The two experiments employ large magnets. Therefore, they are sensitive to ALPs as well as hidden photons (with and without additional MCPs).
CAST, has recently surpassed the HB constraints for ALPs with a two
photon coupling~\cite{Andriamonje:2007ew} (cf. Fig.~\ref{fig:alps_astro}), and its results have been
used to limit a possible solar $\gamma^\prime$ flux~\cite{Redondo:2008aa,Gninenko:2008pz}.
This is shown in Fig.~\ref{fig:hp_astro} as part of the purple area.
Further dedicated helioscopes, an add on to SUMICO~\cite{Ohta:2009un} and a stand alone hidden photon helioscope SHIPS at the Hamburg Observatory~\cite{SHIPSa} are likely to increase the sensitivity for hidden photons towards smaller masses.

To conclude this subsection, the LSW limits on ALPs and MCPs, cf. Figs.~\ref{fig:LSWresult} (a), (b) and (d),
are the most stringent laboratory bounds in the sub-eV mass range. However, for ALPS,
they are still weaker, by nearly three orders of magnitude, in comparison
to the strong limits established by stellar evolution and helioscope searches (cf. Fig.~\ref{fig:alps_astro}).
For MCPs, the situation is even worse (cf. Fig.~\ref{fig:mcp_astro}).
However, such constraints can be relaxed in certain WISP models, in which environmental effects like the temperature
or high density influence the effective photon-WISP coupling. This renders the constraints from the current
generation of LSW experiments still relevant. For hidden photons, the situation is quite different:
in the meV mass range, LSW experiments are already superior to bounds arising from limits on the solar lifetime
and from  CAST  (cf. Fig.~\ref{fig:hp_astro}).

\subsection{Bounds from Big Bang Nucleosynthesis}

Big Bang Nucleosynthesis (BBN) provides us with a unique probe of
the early universe (for a recent review, see
Ref.~\cite{Iocco:2008va}).
The rates of weak and nuclear reactions depend to some extent on the rate of cosmic expansion $H$,
which is proportional to the square root of the energy density $\rho$ of all particles in
the primordial plasma. The larger $\rho$, the less effective the reactions are, modifying the final yield
of the different nuclear species.
For instance, it is well known that the abundance of $^4$He depends crucially on the
freeze-out temperature of the weak interactions that interconvert protons and neutrons,
which depends upon $H$. The extra radiation density $\rho_x$ is normally parametrized with the
effective number of extra thermal neutrino species,
\begin{equation}
N_{\nu,x}^{\rm eff}\equiv  \frac{4}{7}\frac{30}{\pi^2 T^4}\rho_x .
\end{equation}

The success of standard BBN reflected in, among other things, the estimate of this parameter to be
compatible with zero supporting the absence of extra particles present during BBN. For instance,
Ref.~\cite{Simha:2008zj} estimated recently $N_{\nu,x}^{\rm eff} = -0.6_{-0.8}^{+0.9}$ (95\% C.L.) for three standard neutrinos.
However, even more recently, a careful reexamination of the primordial $^4$He abundance with a more detailed
account of systematics has pointed to a higher value, $N_{\nu,x}^{\rm eff} = 0.68_{-0.7}^{+0.8}$ (95\% C.L.),
i.e. on the boundary of implying new physics~\cite{Izotov:2010ca}.

Therefore, while an extra neutral spin-zero particle thermalised during BBN ($N_x=4/7$) is not only allowed
but even preferred by current data,  a spin-zero MCP ($N_x=8/7$) or a massive hidden photon ($N_x=12/7$)
are only marginally allowed, other WISPs like a Dirac mini-charged ($N_x=2$) particle could still be excluded at 95\% C.L.
Thus, the interactions of these MCPs with the standard bath should not allow thermalization before BBN
which leads to a bound $Q < 2\times 10^{-9}$~\cite{Davidson:2000hf} (cf. Fig.~\ref{fig:mcp_astro}, labelled ``BBN").

\subsection{Bounds from the cosmic microwave background\label{Sec:bounds_cmb}}

The cosmic microwave background (CMB) features an almost perfect
blackbody spectrum with ${\cal O}(10^{-5})$ angular anisotropies. It
is released at a temperature $T\sim 0.1$ eV, but the reactions
responsible for the blackbody shape freeze out much earlier, at
$T\sim$ keV. Reactions like $\gamma+...\to$WISP$+...$ would have
depleted photons in a frequency dependent way, which can be
constrained by the precise COBE/FIRAS spectrum
measurements~\cite{Fixsen:1996nj}. This can be used to constrain
light MCPs and ALPs~\cite{Melchiorri:2007sq} as well as hidden
photons~\cite{Jaeckel:2008fi}. More generally~\cite{Mirizzi:2009iz},
(resonant) production of hidden photons leads to distortions in the
CMB spectrum measured by FIRAS strongly constraining their existence
in a wide mass range, as can be seen from Fig.~\ref{fig:hp_astro}
(similar bounds can be obtained for ALPs but they depend on the
unknown strength of the intergalactic magnetic field~
\cite{Mirizzi:2009nq}). Similarly, in presence of MCPs, when the CMB
photons pass through the magnetic field of clusters this leads to a
local distortions of the CMB spectrum in the direction of the
cluster. Such distortions are constrained by
measurements of the so-called Sunaev-Zel'dovich (SZ)
effect and lead to strong bounds on
MCPs~\cite{Burrage:2009yz}\footnote[16]{\base Analogously light from distant
supernovae passing through the (less well known) intergalactic magnetic field
would be dimmed by MCP production, again constraining the existence
of such particles~\cite{Ahlers:2009kh} (SN dimming in
Fig.~\ref{fig:mcp_astro}).}.

On the other hand, around $T\sim$ eV
the primordial plasma is so sparse that WISPs would free-stream out
of the density fluctuations, diminishing their contrast. Moreover,
thermal WISPs contribute to the \emph{radiation} energy density,
delaying the matter-radiation equality and reducing the contrast
growth before decoupling. In this respect, they behave identically
to standard neutrinos~\cite{Ichikawa:2008pz}. Therefore, the extra
contribution to the energy density, $\rho_x$ (and the couplings that
would produce it), can again be constrained from the value of
$N_\nu^{\rm eff}$ inferred from analysis of CMB anisotropies and
other large scale structure (LSS) data, e.g. from a recent
analysis~\cite{Simha:2008zj},
\be
N^{\rm eff}_{\nu,x} =  -0.1^{+2.0}_{-1.4} .
\label{cmb_lss_bound}
\ee
This argument has been used to constrain ALPs~\cite{Hannestad:2010yi}
and meV $\gamma^\prime$s~\cite{Jaeckel:2008fi}
(cf. Fig.~\ref{fig:hp_astro}, labelled ``FIRAS+hCMB").

Interestingly, some global cosmological analyses
that take into account precision cosmological
data on the cosmic microwave background and on the large scale structure of the universe
appear to require some extra radiation energy density from invisible particles apart from the three known neutrino
species.
The case for this was strengthened by the recently released WMAP 7 year data, whose
global analysis points to a value of the effective
number of neutrinos higher than the standard value of three by an amount
$\Delta N_{\nu}^{\rm eff}=1.3\pm 0.9$~\cite{Komatsu:2010fb}.
Hidden photons in the parameter region indicated by a band in Fig.~\ref{fig:LSWresult} (c)
would lead to a natural explanation of this finding~\cite{Jaeckel:2008fi}.
However, the new data from ALPS excludes this possibility nearly entirely, up to
a tiny region in parameter space, $m_{\gamma^\prime}\approx 0.18$~meV, $\chi\approx 1.4\times 10^{-6}$.

\subsection{Hints for cosmic photon regeneration\label{Sec:hints_cosmic}}

It has been argued that recent observations in TeV gamma astronomy may point towards the
existence of ALPs with a very small mass,
\begin{equation}
m_{\phi}\ll 10^{-9}\,{\rm eV},
\label{mass_benchmark_ALP_transp}
\end{equation}
and a photon coupling in the range
\begin{equation}
g\sim 10^{-12}\div 10^{-11}\ {\rm GeV}^{-1}.
\label{g_benchmark_ALP_transp}
\end{equation}
Quite distant astrophysical sources have been observed by Cherenkov telescopes like H.E.S.S. and
MAGIC. This appeared to be quite puzzling, since the gamma ray absorption rate
due to electron/positron pair production off the extragalactic background light (EBL)
was believed to be too strong to allow
for their observation~\cite{Aharonian:2005gh,Mazin:2007pn}.
Clearly, a conventional explanation is either that the EBL is less dense than expected and/or that the source
spectra are harder than previously thought.
Alternatively, such a high transparency of the universe may also be explained by cosmic LSW: the conversion of
gamma rays into ALPs in the magnetic fields around the gamma ray sources or in the intergalactic medium, followed by their unimpeded travel towards our galaxy and the consequent reconversion into photons in the galactic/intergalactic magnetic
fields~\cite{Hochmuth:2007hk,Hooper:2007bq,DeAngelis:2007dy}.
The intergalactic magnetic fields are not well known but the assumption of being organized in randomly oriented patches would produce a relevant dispersion of the photon transfer function around the mean value.
A powerful signature of cosmic photon regeneration could therefore emerge if the reconstructed
EBL along different lines of sight towards different TeV gamma sources were to display such a
characteristic scatter~\cite{Mirizzi:2009aj}.
However, to accomplish that much more data from many more quite distant TeV gamma sources along different directions in the sky has to be collected.
It would be great, if we were able to probe the same range of parameters
even earlier in the laboratory, by laser light shining through a wall.

\section{The next generation of LSW\label{sec:next}}

Currently, the next generation of LSW experiments is being developed. Two targets in WISP parameter
space have been identified upon which this next generation should shoot~\cite{Ringwald:2010yr,Baker:2010ma}.
They arise immediately from
the astrophysical and cosmological observations discussed in the last section and constitute
both
\begin{itemize}
\item {\bf challenges} to increase the sensitivity beyond astro, cosmo, and other lab bounds, and
\item {\bf opportunities} to test the WISP interpretation of hints for cosmic photon regeneration.
\end{itemize}

Fortunately, this seems doable.
The current state-of-the-art LSW experiment,
ALPS, exploiting an optical resonator, with a power-build up of $\beta_g\sim 300$ at the generation side
of the experiment, resulting in a power of $\beta_g {\cal P}_{\rm prim}\sim 1.2$~kW available for $\gamma\to$~WISP conversions,
established an upper limit $P_{\rm LSW}\lesssim {\rm few}\times 10^{-25}$ on the LSW probability~\cite{Ehret:2010mh}.
Now, exploiting additionally a high finesse
($\beta_r\sim 10^4$) optical resonator also on the regeneration side of the
experiment~\cite{Hoogeveen:1990vq,Sikivie:2007qm} and a single-photon
counter, together with an increased power buildup, by a factor of $\sim 100$, on the generation side,
it seems possible to improve the sensitivity on the LSW probability by $\sim 4+2+2 = 8$ orders of magnitude,
corresponding to an improved sensitivity on the ALP coupling  $g$ or hidden photon coupling
$\chi$ by $\sim 8/4 = 2$ orders of magnitude.

\begin{figure}[t!]
\centering
\includegraphics[width=.5\textwidth]{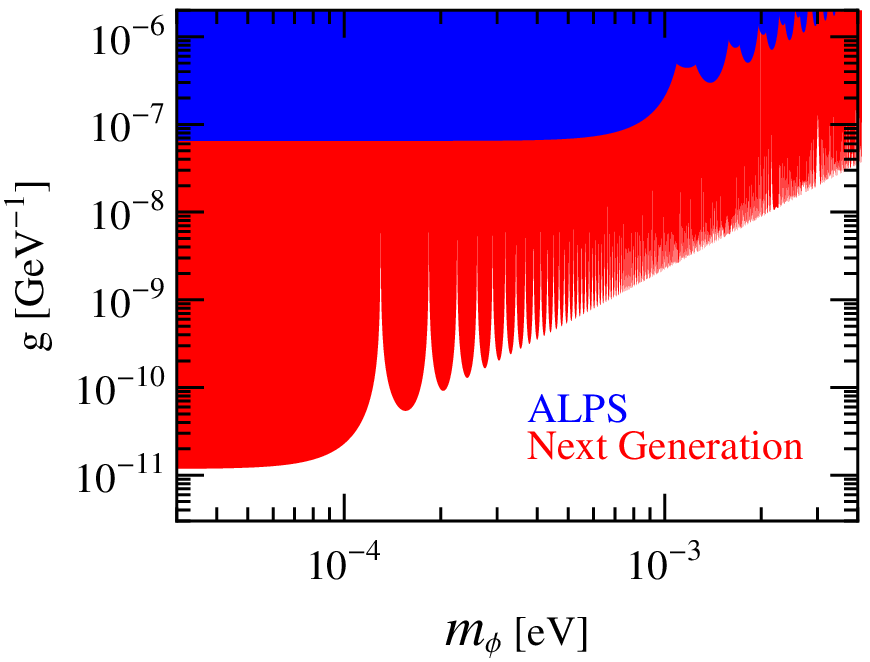}

\includegraphics[width=.4\textwidth]{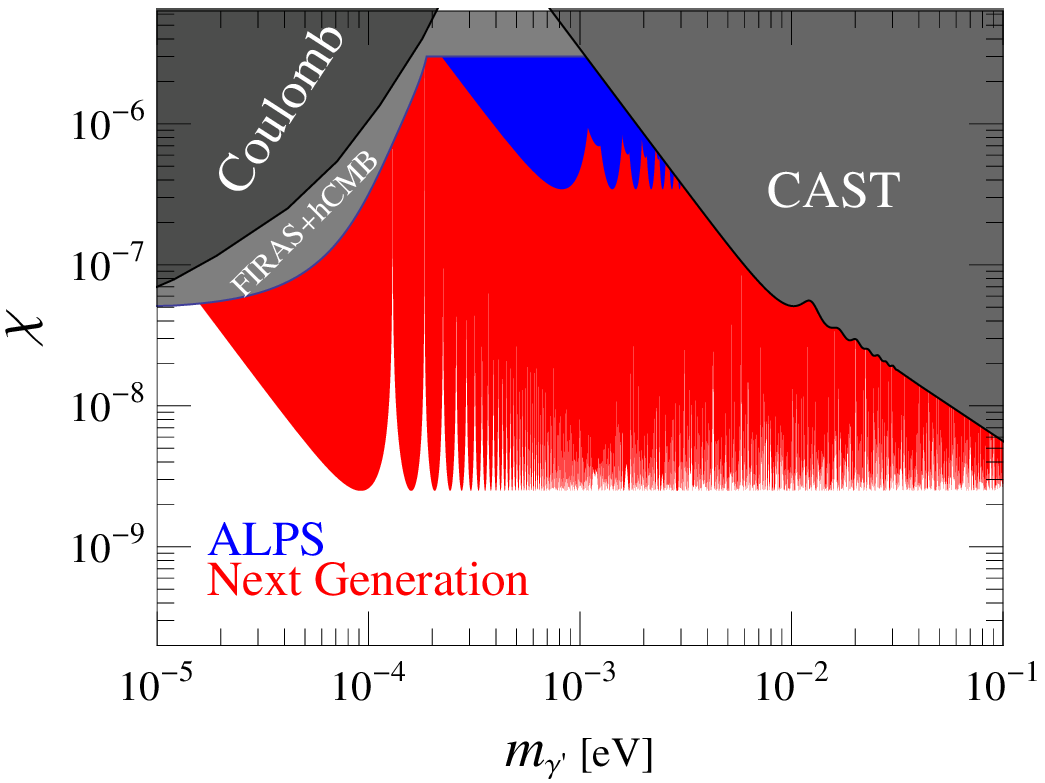} 
\includegraphics[width=.4\textwidth]{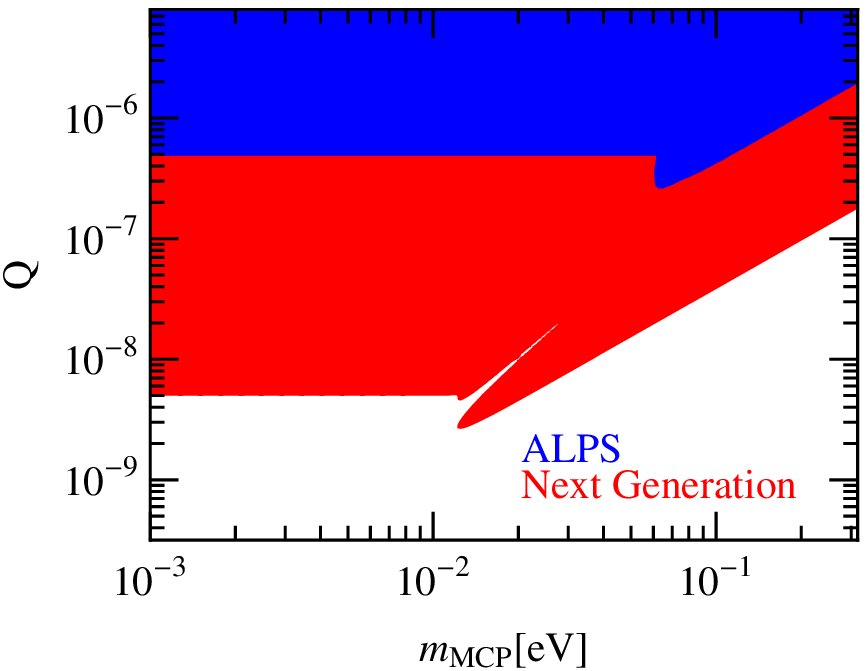}
\caption{
Projected WISP sensitivity of a next generation LSW experiments (red) compared with the current experimental bounds  (blue) (from the results of~\cite{Arias:2010bh}).
The benchmark for the next generation is taken to be a 20+20 HERA magnet configuration
with $300$~kW of $1.17$~eV photons in the generation part and a power buildup $\beta_r= 10^{5}$ on the regeneration side
and a single-photon counter with dark current $\sim 10^{-4}$ counts/sec,
corresponding to a projected sensitivity of $\sim 10^{-33}$ on the LSW probability.
}
\label{fig:LSWnext}
\end{figure}

In the ALP case, there are further improvements possible:
increasing $B\times L$, the magnitude times the length of the magnetic field region,
by more than one order of magnitude compared to the current experiments,
e.g. by exploiting 20+20 HERA magnets~\cite{Ringwald:2003nsa} at ALPS, instead of the current 1/2+1/2 configuration.
one may reach a sensitivity in the $g\sim {\rm few}\times 10^{-11}$~GeV$^{-1}$ range, for
light ALPs, $m_\phi\ll$~meV~\cite{Mueller:2009wt,Arias:2010bh}.
We display in Fig.~\ref{fig:LSWnext} the sensitivity of a realistic setup along these lines for ALPs, HPs and
MCPs. Obviously, for ALPs and HPs, the above mentioned challenge is met: the next generation of
LSW experiments will have a hugely enlarged discovery potential, since its sensitivity exceeds the
stellar, solar and cosmological bounds.
Clearly, an even wider range of opportunities for discovery would open up if the sensitivity in $g$
can be improved even more, by one order of magnitude, down to $g\sim {\rm few}\times 10^{-12}$~GeV$^{-1}$,
possibly by a combination of laser and magnet upgrades.

Unfortunately, with the present ideas it seems impossible to probe the QCD axion itself in
purely laboratory based LSW experiments. This is because coherence is lost already at
masses well below the QCD axion mass. To counteract this effect\footnote[17]{There have been other proposals such
as exploiting alternating directions of the magnetic field~\cite{VanBibber:1987rq},
phase shift plates~\cite{Jaeckel:2007gk}, and gaps~\cite{Arias:2010bh} between the
magnets, but all of them seem not sufficient to close the gap in mass up to the QCD axion.},
one may contemplate about
LSW with more energetic photons, e.g. from synchrotron
radiation sources or X-ray free-electron lasers~\cite{Rabadan:2005dm,Dias:2009ph}
(for a pioneering experiment in this direction see Ref.~\cite{Battesti:2010dm} and Table~\ref{tab:LSWexp}).
However, currently this  does not seem promising
because the average photon flux of current or planned facilities is just in the $10^{17}$ photons
per second range, corresponding to an average power in the $10^{-2}$~W ballpark, many orders
of magnitude smaller than the one achievable in optical resonant cavities ($\sim 100$ kW).

\section{Conclusions\label{sec:conclusions}}

Considerable activity takes place presently in the field of laser light shining through a wall experiments,
which is now preparing a new generation.
Important advances in laser technology appear to pave the way to beat the sensitivity of current
ALP and other WISP helioscopes
and to probe explanations of astrophysical puzzles in terms of photon $\leftrightarrow$ WISP oscillations.
Pioneering experiments exploiting instead high-quality microwave cavities for the generation and regeneration of
WISPs are in the commissioning phase. The microwave cavity search for dark matter axions probes a complementary
region in parameter space compared to the other photon regeneration experiments and should provide a definitive
answer whether QCD axions are the dominant part of cold dark matter within the current
decade~\cite{Duffy:2006aa,Asztalos:2009yp}.

Clearly, the detection of a WISP in an LSW experiment
will deeply change our view of cosmology and astrophysics. In fact, this reminds us on the neutrino story: they were postulated and confirmed as subtle effects in laboratory experiments and nowadays
they are essential ingredients in our understanding of astrophysics
(e.g. in white dwarf cooling or supernova type-II explosions) and cosmology (e.g. in big bang nucleosynthesis
or structure formation). Furthermore, WISPs have been recently shown to have possibly interesting technological applications, such as for example large distance communications through matter~\cite{Stancil:2007yk,Jaeckel:2009wm}.

All in all, LSW experiments may give important information about fundamental particle
physics complementary to the one obtainable at high energy colliders. Already today these
experiments provide very strong bounds on light weakly interacting particles. But even
more excitingly the next decade is likely to bring considerable advances and huge
discovery potential for new physics.

\section*{Acknowledgements}

The authors would like to thank Markus Ahlers, Paola Arias, Clare Burrage, Holger Gies,
Mark Goodsell, Joerg Jaeckel, Axel Lindner, and Alessandro Mirizzi for interesting discussions, helpful suggestions and joyful collaboration on the subject discussed in this review. J.R. acknowledges support from
the SFB 676 and the DFG cluster of excellence EXC 153 Origin and Structure of the
Universe.


\section*{About the authors}

\noindent
Javier Redondo is a Postdoctoral Researcher at the Max-Planck-Institute
for Physics, Munich, Germany. He received his Ph.D. in physics
from the Universitat Autonoma de Barcelona, Spain, for his work
on the particle interpretation of the PVLAS anomaly and the question
whether it can be compatible with astrophysical bounds. As a
Postdoctoral Researcher in the Theory Group at DESY he joined
the ALPS experiment, where he focused on the development of the
theoretical implications of the (non-)observation of LSW.
Furthermore, he initiated the SHIPS helioscope which is currently being setup
at the Hamburg Observatory.
\\

\noindent
Andreas Ringwald is Staff Member in the Theory Group at
DESY, Hamburg, Germany. He received his Ph.D. in physics
from the University of Heidelberg. His research has
centred on theoretical particle physics, astro-particle physics, and
particle-cosmology throughout his career. He is very much interested not only to
develop the theory and phenomenology of WISPs but also in 
setting up laboratory experiments searching for them.
He initiated the experimental programme at DESY for ALPs (and other WISPs) searches
and currently serves as a co-spokesperson of the ALPS experiment.

\label{lastpage}

\end{document}